\documentclass[12pt,a4paper]{article}

\usepackage{ifthen} 
\newboolean{pdflatex}
\setboolean{pdflatex}{true} 

\newboolean{articletitles}
\setboolean{articletitles}{true} 

\newboolean{uprightparticles}
\setboolean{uprightparticles}{false} 

\def\paperauthors{LHCb collaboration} 
\def\paperasciititle{Measurement of prompt charged-particle production in proton-proton collisions at a centre-of-mass energy of 13 TeV} 
\def\papertitle{Measurement of prompt charged-particle production\\in \proton\proton collisions at ${\sqs = 13\tev}$} 
\def\paperkeywords{{High Energy Physics}, {LHCb}} 
\def\papercopyright{\the\year\ CERN for the benefit of the LHCb collaboration} 
\def\paperlicence{CC BY 4.0 licence}
\def\paperlicenceurl{https://creativecommons.org/licenses/by/4.0/}

\usepackage{booktabs}

\usepackage[top=1in, bottom=1.25in, left=1in, right=1in]{geometry}

\usepackage{microtype}
\usepackage{lineno}  
\usepackage{xspace} 
\usepackage{caption} 

\usepackage{graphicx}  
\usepackage{color}
\usepackage{colortbl}
\graphicspath{{./figs/}} 

\usepackage{amsmath} 
\usepackage{amssymb}
\usepackage{amsfonts}
\usepackage{upgreek} 

\newcommand*\patchAmsMathEnvironmentForLineno[1]{%
\expandafter\let\csname old#1\expandafter\endcsname\csname #1\endcsname
\expandafter\let\csname oldend#1\expandafter\endcsname\csname
end#1\endcsname
 \renewenvironment{#1}%
   {\linenomath\csname old#1\endcsname}%
   {\csname oldend#1\endcsname\endlinenomath}%
}
\newcommand*\patchBothAmsMathEnvironmentsForLineno[1]{%
  \patchAmsMathEnvironmentForLineno{#1}%
  \patchAmsMathEnvironmentForLineno{#1*}%
}
\AtBeginDocument{%
\patchBothAmsMathEnvironmentsForLineno{equation}%
\patchBothAmsMathEnvironmentsForLineno{align}%
\patchBothAmsMathEnvironmentsForLineno{flalign}%
\patchBothAmsMathEnvironmentsForLineno{alignat}%
\patchBothAmsMathEnvironmentsForLineno{gather}%
\patchBothAmsMathEnvironmentsForLineno{multline}%
\patchBothAmsMathEnvironmentsForLineno{eqnarray}%
}

\usepackage{hyperxmp}

\usepackage[pdftex,
            pdfauthor={\paperauthors},
            pdftitle={\paperasciititle},
            pdfkeywords={\paperkeywords},
            pdfcopyright={Copyright (C) \papercopyright},
            pdflicenseurl={\paperlicenceurl}]{hyperref}

\usepackage[colorinlistoftodos,textsize=scriptsize]{todonotes}

\usepackage[bottom,flushmargin,hang,multiple]{footmisc}

\usepackage[all]{hypcap} 

\usepackage{xspace} 
\usepackage{upgreek}

\def\lhcb   {\mbox{LHCb}\xspace}
\def\atlas  {\mbox{ATLAS}\xspace}
\def\cms    {\mbox{CMS}\xspace}
\def\alice  {\mbox{ALICE}\xspace}

\def\lhc    {\mbox{LHC}\xspace}

\def\MagUp {\mbox{\em Mag\kern -0.05em Up}\xspace}

\ifthenelse{\boolean{uprightparticles}}%
{

 \def\Pmu         {\ensuremath{\upmu}\xspace}

 \def\Ppi         {\ensuremath{\uppi}\xspace}

 \def\Ppsi        {\ensuremath{\uppsi}\xspace}

 \def\PDelta      {\ensuremath{\Delta}\xspace}                 
 \def\PXi         {\ensuremath{\Xi}\xspace}                 
 \def\PLambda     {\ensuremath{\Lambda}\xspace}                 
 \def\PSigma      {\ensuremath{\Sigma}\xspace}                 
 \def\POmega      {\ensuremath{\Omega}\xspace}                 
 \def\PUpsilon    {\ensuremath{\Upsilon}\xspace}

 \def\PB      {\ensuremath{\mathrm{B}}\xspace}                 
                  
 \def\PD      {\ensuremath{\mathrm{D}}\xspace}

 \def\PJ      {\ensuremath{\mathrm{J}}\xspace}                 
 \def\PK      {\ensuremath{\mathrm{K}}\xspace}

 \def\Pb      {\ensuremath{\mathrm{b}}\xspace}                 
 \def\Pc      {\ensuremath{\mathrm{c}}\xspace}                 
                  
 \def\Pe      {\ensuremath{\mathrm{e}}\xspace}

 \def\Ph      {\ensuremath{\mathrm{h}}\xspace}                 
 \def\Pi      {\ensuremath{\mathrm{i}}\xspace}

 \def\Pp      {\ensuremath{\mathrm{p}}\xspace}

 \def\Ps      {\ensuremath{\mathrm{s}}\xspace}

 \def\thebaroffset{0.0em}
}
{

 \def\Pmu         {\ensuremath{\mu}\xspace}

 \def\Ppi         {\ensuremath{\pi}\xspace}

 \def\Ppsi        {\ensuremath{\psi}\xspace}                 
                  
 \mathchardef\PDelta="7101
 \mathchardef\PXi="7104
 \mathchardef\PLambda="7103
 \mathchardef\PSigma="7106
 \mathchardef\POmega="710A
 \mathchardef\PUpsilon="7107
                  
 \def\PB      {\ensuremath{B}\xspace}                 
                  
 \def\PD      {\ensuremath{D}\xspace}

 \def\PJ      {\ensuremath{J}\xspace}                 
 \def\PK      {\ensuremath{K}\xspace}

 \def\Pb      {\ensuremath{b}\xspace}                 
 \def\Pc      {\ensuremath{c}\xspace}                 
                  
 \def\Pe      {\ensuremath{e}\xspace}

 \def\Ph      {\ensuremath{h}\xspace}                 
 \def\Pi      {\ensuremath{i}\xspace}

 \def\Pp      {\ensuremath{p}\xspace}

 \def\Ps      {\ensuremath{s}\xspace}

 \def\thebaroffset{0.18em}
}
\newcommand{\offsetoverline}[2][\thebaroffset]{\kern #1\overline{\kern -#1 #2}}%

\makeatletter
\ifcase \@ptsize \relax
  \newcommand{\miniscule}{\@setfontsize\miniscule{4}{5}}
\or
  \newcommand{\miniscule}{\@setfontsize\miniscule{5}{6}}
\or
  \newcommand{\miniscule}{\@setfontsize\miniscule{5}{6}}
\fi
\makeatother

\DeclareRobustCommand{\optbar}[1]{\shortstack{{\miniscule (\rule[.5ex]{1.25em}{.18mm})}
  \\ [-.7ex] $#1$}}

\def\en         {{\ensuremath{\Pe^-}}\xspace}   

\def\mun        {{\ensuremath{\Pmu^-}}\xspace} 

\def\mumu       {{\ensuremath{\Pmu^+\Pmu^-}}\xspace}

\def\squark    {{\ensuremath{\Ps}}\xspace}

\def\cquark    {{\ensuremath{\Pc}}\xspace}

\def\bquark    {{\ensuremath{\Pb}}\xspace}

\def\hadron {{\ensuremath{\Ph}}\xspace}
\def\pion   {{\ensuremath{\Ppi}}\xspace}

\def\pip    {{\ensuremath{\pion^+}}\xspace}
\def\pim    {{\ensuremath{\pion^-}}\xspace}

\def\kaon    {{\ensuremath{\PK}}\xspace}

\def\KorKbar {\kern \thebaroffset\optbar{\kern -\thebaroffset \PK}{}\xspace}

\def\Kp      {{\ensuremath{\kaon^+}}\xspace}
\def\Km      {{\ensuremath{\kaon^-}}\xspace}

\def\KS      {{\ensuremath{\kaon^0_{\mathrm{S}}}}\xspace}
\def\Vzero   {{\ensuremath{V^0}}\xspace}

\def\D       {{\ensuremath{\PD}}\xspace}

\def\DorDbar {\kern \thebaroffset\optbar{\kern -\thebaroffset \PD}\xspace}

\def\Dp      {{\ensuremath{\D^+}}\xspace}
\def\Dm      {{\ensuremath{\D^-}}\xspace}

\def\DpDm    {\ensuremath{\Dp {\kern -0.16em \Dm}}\xspace}

\def\B       {{\ensuremath{\PB}}\xspace}

\def\BorBbar {\kern \thebaroffset\optbar{\kern -\thebaroffset \PB}\xspace}

\def\Bd      {{\ensuremath{\B^0}}\xspace}

\def\BdorBdbar {\kern \thebaroffset\optbar{\kern -\thebaroffset \Bd}\xspace}

\def\Bs      {{\ensuremath{\B^0_\squark}}\xspace}

\def\BsorBsbar {\kern \thebaroffset\optbar{\kern -\thebaroffset \Bs}\xspace}

\def\jpsi     {{\ensuremath{{\PJ\mskip -3mu/\mskip -2mu\Ppsi}}}\xspace}

\def\Y#1S{\ensuremath{\PUpsilon{(#1S)}}\xspace}

\def\proton      {{\ensuremath{\Pp}}\xspace}
\def\antiproton  {{\ensuremath{\overline \proton}}\xspace}

\def\Lz          {{\ensuremath{\PLambda}}\xspace}
\def\Lbar        {{\ensuremath{\offsetoverline{\PLambda}}}\xspace}
\def\LorLbar     {\kern \thebaroffset\optbar{\kern -\thebaroffset \PLambda}\xspace}

\def\Sigmares    {{\ensuremath{\PSigma}}\xspace}

\def\Sigmap      {{\ensuremath{\Sigmares{}^+}}\xspace}
\def\Sigmam      {{\ensuremath{\Sigmares{}^-}}\xspace}

\def\Xires       {{\ensuremath{\PXi}}\xspace}

\def\Xim         {{\ensuremath{\Xires^-}}\xspace}

\def\Omegares    {{\ensuremath{\POmega}}\xspace}

\def\Omegam      {{\ensuremath{\Omegares^-}}\xspace}

\newcommand{\decay}[2]{\ensuremath{#1\!\to #2}\xspace} 

\def\to                 {\ensuremath{\rightarrow}\xspace}

\def\eps   {{\ensuremath{\varepsilon}}\xspace}

\def\AT#1     {\ensuremath{A_{\mathrm{T}}^{#1}}\xspace}           

\def\C#1      {\ensuremath{\mathcal{C}_{#1}}\xspace}                       
\def\Cp#1     {\ensuremath{\mathcal{C}_{#1}^{'}}\xspace}                    
\def\Ceff#1   {\ensuremath{\mathcal{C}_{#1}^{\mathrm{(eff)}}}\xspace}        
\def\Cpeff#1  {\ensuremath{\mathcal{C}_{#1}^{'\mathrm{(eff)}}}\xspace}       
\def\Ope#1    {\ensuremath{\mathcal{O}_{#1}}\xspace}                       
\def\Opep#1   {\ensuremath{\mathcal{O}_{#1}^{'}}\xspace}                    

\newcommand{\nospaceunit}[1]{\ensuremath{\text{#1}}}       
\newcommand{\aunit}[1]{\ensuremath{\text{\,#1}}}       

\newcommand{\tev}{\aunit{Te\kern -0.1em V}\xspace}
\newcommand{\gev}{\aunit{Ge\kern -0.1em V}\xspace}
\newcommand{\mev}{\aunit{Me\kern -0.1em V}\xspace}
\newcommand{\kev}{\aunit{ke\kern -0.1em V}\xspace}
\newcommand{\ev}{\aunit{e\kern -0.1em V}\xspace}
 
\newcommand{\mevc}{\ensuremath{\aunit{Me\kern -0.1em V\!/}c}\xspace}
\newcommand{\gevc}{\ensuremath{\aunit{Ge\kern -0.1em V\!/}c}\xspace}
\newcommand{\mevcc}{\ensuremath{\aunit{Me\kern -0.1em V\!/}c^2}\xspace}
\newcommand{\gevcc}{\ensuremath{\aunit{Ge\kern -0.1em V\!/}c^2}\xspace}

\def\mm   {\aunit{mm}\xspace}
\def\mma  {\ensuremath{\aunit{mm}^2}\xspace}
\def\mum  {\ensuremath{\,\upmu\nospaceunit{m}}\xspace}

\def\mbarn{\aunit{mb}\xspace}

\def\nb {\aunit{nb}\xspace}
\def\invnb {\ensuremath{\nb^{-1}}\xspace}

\def\ns   {\ensuremath{\aunit{ns}}\xspace}
\def\ps   {\ensuremath{\aunit{ps}}\xspace}

\def\deriv {\ensuremath{\mathrm{d}}}

\def\gsim{{~\raise.15em\hbox{$>$}\kern-.85em
          \lower.35em\hbox{$\sim$}~}\xspace}
\def\lsim{{~\raise.15em\hbox{$<$}\kern-.85em
          \lower.35em\hbox{$\sim$}~}\xspace}

\def\sqs   {\ensuremath{\protect\sqrt{s}}\xspace}

\def\pt         {\ensuremath{p_{\mathrm{T}}}\xspace}

\def\ptot       {\ensuremath{p}\xspace}

\newcommand{\lum} {\ensuremath{\mathcal{L}}\xspace}

\def\evtgen     {\mbox{\textsc{EvtGen}}\xspace}

\def\geant      {\mbox{\textsc{Geant4}}\xspace}

\def\photos     {\mbox{\textsc{Photos}}\xspace}

\def\pythia     {\mbox{\textsc{Pythia}}\xspace}

\def\tell1  {TELL1\xspace}
\def\ukl1   {UKL1\xspace}

\newcommand{\eg}{\mbox{\itshape e.g.}\xspace}
\newcommand{\ie}{\mbox{\itshape i.e.}\xspace}

\usepackage{cite} 
\usepackage{mciteplus}

\usepackage{longtable} 

\begin{document}

\renewcommand{\thefootnote}{\fnsymbol{footnote}}
\setcounter{footnote}{1}


\begin{titlepage}
\pagenumbering{roman}

\vspace*{-1.5cm}
\centerline{\large EUROPEAN ORGANIZATION FOR NUCLEAR RESEARCH (CERN)}
\vspace*{1.5cm}
\noindent
\begin{tabular*}{\linewidth}{lc@{\extracolsep{\fill}}r@{\extracolsep{0pt}}}
\ifthenelse{\boolean{pdflatex}}
{\vspace*{-1.5cm}\mbox{\!\!\!\includegraphics[width=.14\textwidth]{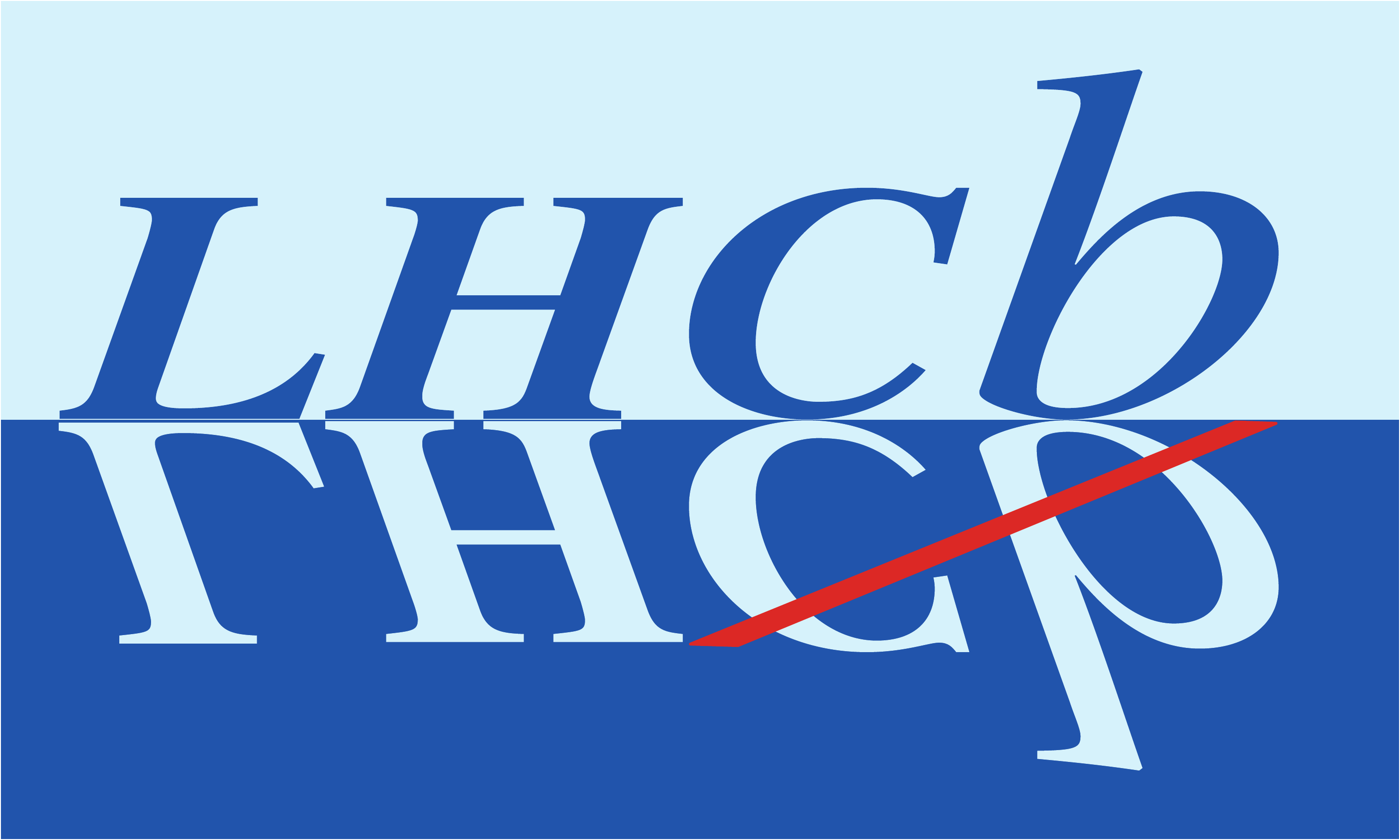}} & &}%
{\vspace*{-1.2cm}\mbox{\!\!\!\includegraphics[width=.12\textwidth]{figs/lhcb-logo.eps}} & &}%
\\
 & & CERN-EP-2021-110 \\  
 & & LHCb-PAPER-2021-010 \\  
 & & 4 April 2022 \\ 
 & & \\
\end{tabular*}

\vspace*{4.0cm}

{\normalfont\bfseries\boldmath\huge
\begin{center}
  \papertitle
\end{center}
}

\vspace*{2.0cm}

\begin{center}
\paperauthors\footnote{Authors are listed at the end of this paper.}
\end{center}

\vspace{\fill}

\begin{abstract}
  \noindent
  The differential cross-section of prompt inclusive production of long-lived charged particles in proton-proton collisions is measured using a data sample recorded by the \lhcb experiment at a centre-of-mass energy of ${\sqs = 13\tev}$. The data sample, collected with an unbiased trigger, corresponds to an integrated luminosity of $5.4\invnb$. The differential cross-section is measured as a function of transverse momentum and pseudorapidity in the ranges ${\pt \in [80, 10\,000)\mevc}$ and ${\eta \in [2.0, 4.8)}$ and is determined separately for positively and negatively charged particles. The results are compared with predictions from various hadronic-interaction models.
\end{abstract}

\vspace*{2.0cm}

\begin{center}
  Published in JHEP 01 (2022) 166
\end{center}

\vspace{\fill}

{\footnotesize
\centerline{\copyright~\papercopyright. \href{\paperlicenceurl}{\paperlicence}.}}
\vspace*{2mm}

\end{titlepage}


\newpage
\setcounter{page}{2}
\mbox{~}
%
%
%
%


\renewcommand{\thefootnote}{\arabic{footnote}}
\setcounter{footnote}{0}

\cleardoublepage


\pagestyle{plain} 
\setcounter{page}{1}
\pagenumbering{arabic}


\section{Introduction}
\label{sec:introduction}

Hadron production in inelastic high-energy proton-proton (\proton\proton) collisions is dominated by soft processes described by quantum chromodynamics (QCD). These processes cannot be calculated from first principles in perturbative QCD due to the large coupling constant $\alpha_\mathrm{s}$ at small average momentum transfer. Instead, predictions are based on phenomenological models. The determination of the model parameters relies on input from experiments at particle accelerators. Monte Carlo event generators, in which these models are implemented, are used at the Large Hadron Collider (\lhc) to simulate the final-state particles originating from the soft component of a collision. An introduction to soft-QCD theories is presented, for example, in Refs.~\cite{Low:1975sv,Mueller:1986ey,Nikolaev:1991et,Goulianos:1994ph,Drescher:2000ha}.

In the field of cosmic-ray research, generators are employed to simulate interactions of ultra-relativistic nuclei with the atmosphere of the Earth, which induce extensive particle cascades, referred to as air showers. Although often used to predict interactions in a phase space that is not covered by the input from experiments, the generators are remarkably successful at describing many features of air showers. However, a long-standing excess is observed in the number of muons produced in high-energy air showers compared to simulations, termed the Muon Puzzle~\cite{Dembinski:2019uta,Dembinski:2020flw}. Precision measurements of the production of light hadrons at the ${\!\tev}$ scale in the forward region are needed to guide further and constrain the models~\cite{Citron:2018lsq,Baur:2019cpv,Albrecht:2021yla} and to extrapolate them safely to even higher collision energies that are of interest in astroparticle physics. The \lhcb detector covers the forward pseudorapidity range ${2 < \eta < 5}$, which is of particular interest for cosmic-ray research.

A suitable proxy for the production of light hadrons is the production of prompt long-lived charged particles. A particle is classified as long-lived if its lifetime is greater than $30\ps$, and prompt if it is either produced directly in the primary interaction or does not have long-lived ancestors~\cite{Acharya:2017a}. Long-lived charged particles are electrons, muons, pions, kaons and protons as well as \Sigmap, \Sigmam, \Xim and \Omegam baryons and their antiparticles.

Approaches based on Gribov-Regge field theory model soft QCD processes via soft and hard Pomeron interactions. In the approach used by \textsc{Sibyll}~2.3d~\cite{Engel:2019dsg} and \textsc{DPMJet}~III~\cite{Roesler:2000he,Fedynitch:2015kcn}, the two regimes are rather decoupled and the pseudorapidity distribution of prompt long-lived charged particles is narrow, while the unified approach used by \textsc{EPOS-LHC}~\cite{Pierog:2013ria} and \textsc{QGSJet}~II-04~\cite{Ostapchenko:2010vb} produces wider distributions~\cite{Albrecht:2021yla}. The generator \pythia~\cite{Sjostrand:2014zea,*Sjostrand:2006za}, mentioned for completeness, is a parton-shower model that follows a different approach.

Measurements of charged-particle spectra over a wide pseudorapidity range make it possible to discriminate between these approaches. Charged-particle spectra in \proton\proton collisions have been studied as a function of pseudorapidity, $\eta$, or transverse momentum, \pt, by the \alice (${|\eta| < 2}$)~\cite{ALICE:2015qqj,ALICE:2018vuu}, \atlas (${|\eta| < 2.5}$)~\cite{ATLAS:2010jvh,ATLAS:2016qux,ATLAS:2016zkp,ATLAS:2016zba}, \cms (${|\eta| < 2.2}$)~\cite{CMS:2011mry,CMS:2014kix,CMS:2015zrm,CMS:2017dou,CMS:2017eoq,CMS:2018nhd}, TOTEM (${5.3 < |\eta| < 6.5}$)~\cite{CMS:2014kix} and \lhcb (${2.0 < |\eta| < 4.8}$)~\cite{LHCb-PAPER-2013-070} collaborations. Spectra of identified pions, kaons, and protons and other hadrons of light flavour have been measured by the \alice (${|\eta| < 0.5}$)~\cite{ALICE:2015ial,ALICE:2020jsh} and \cms (${|\eta| < 1}$)~\cite{CMS:2017eoq} collaborations. The \lhcb collaboration studied ratios of these particle species (${2.5 < |\eta| < 4.5}$)~\cite{LHCb-PAPER-2011-037}, while the LHCf collaboration studied very forward spectra (${|\eta| > 8.4}$) of neutral pions~\cite{LHCf:2012mtr,LHCf:2015rcj} and neutrons~\cite{LHCf:2015nel,LHCf:2018gbv,LHCf:2020hjf}. The inclusive energy spectrum, which includes charged and neutral particles that are overwhelmingly of hadronic origin, was measured by the \cms (${5.2 < |\eta| < 6.6}$)~\cite{CMS:2017dou} and \lhcb (${1.9 < |\eta| < 4.9}$)~\cite{LHCb-PAPER-2012-034} collaborations.

In this paper, a measurement of the double-differential cross-section of inclusive production of prompt long-lived charged particles in \proton\proton collisions at a centre-of-mass energy of ${\sqs = 13\tev}$ is presented. The data sample was recorded by the \lhcb experiment with a zero-bias trigger in 2015 and corresponds to an integrated luminosity of $5.4\invnb$. The double-differential cross-section is measured separately for positively and negatively charged particles as a function of transverse momentum and pseudorapidity in intervals that span the ranges ${\pt \in [80, 10\,000)\mevc}$ and ${\eta \in [2.0, 4.8)}$. The minimum transverse momentum is a function of $\eta$ due to the fiducial acceptance of the spectrometer. Both the charge-combined differential cross-section and the ratio of the differential cross-sections for the two charges are compared with predictions from four different hadronic-interaction models. The measurement goes a step further compared to the precursors in several ways. It is a double-differential measurement in the forward region performed separately for each charge, non-prompt production is removed and there is no trigger bias. High precision is achieved thanks to the high-performance tracker of the \lhcb experiment and the detailed control measurements that can be performed with a general-purpose spectrometer.

The paper is structured as follows. In Sect.~\ref{sec:detector}, the detector as well as the data and simulated samples used in this measurement are described. The analysis strategy is presented in Sect.~\ref{sec:analysis_strategy}. The efficiencies and the background contributions are detailed in Sects.~\ref{sec:efficiencies} and~\ref{sec:background_contributions}, respectively. In Sect.~\ref{sec:results}, the results are discussed, and a summary is provided in Sect.~\ref{sec:summary}.

\section{Detector and data sample}
\label{sec:detector}

The \lhcb detector~\cite{LHCb-DP-2008-001,LHCb-DP-2014-002} is a single-arm forward spectrometer covering the \mbox{pseudorapidity} range $2<\eta <5$, designed for the study of particles containing \bquark or \cquark quarks. The detector includes a high-precision tracking system consisting of a silicon-strip vertex detector surrounding the $pp$ interaction region, a large-area silicon-strip detector located upstream of a dipole magnet with a bending power of about $4{\mathrm{\,Tm}}$, and three stations of silicon-strip detectors and straw drift tubes placed downstream of the magnet. The tracking system provides a measurement of the momentum, \ptot, of charged particles with a relative uncertainty that varies from 0.5\% at low momentum to 1.0\% at 200\gevc. The minimum distance of a track to a primary $pp$ collision vertex, the impact parameter, is measured with a resolution of $(15+29/\pt)\mum$, where \pt is the component of the momentum transverse to the beam, in\,\gevc. Different types of charged hadrons are distinguished using information from two ring-imaging Cherenkov detectors. Photons, electrons and hadrons are identified by a calorimeter system consisting of scintillating-pad and preshower detectors, an electromagnetic and a hadronic calorimeter. Muons are identified by a system composed of alternating layers of iron and multiwire proportional chambers.

The online event selection for this measurement is performed by an unbiased trigger. Therefore, no trigger-related systematic uncertainty arises. At the hardware stage, events are accepted at a fixed rate. The software stage then restricts the data sample to collisions of leading bunches of the \lhc bunch trains, which avoids background from previous events, while the bunch spacing of $50\ns$ in these low-intensity runs avoids contributions to the read-out from following events. The analysed data sample contains the events from two \lhc fills, recorded with opposite magnetic-field configurations of the \lhcb dipole magnet. The field configuration that bends positively (negatively) charged particles in the horizontal plane towards the centre of the \lhc ring is referred to as upwards (downwards). The fill recorded with the magnetic field pointing upwards comprises ${226 \times 10^6}$ events and corresponds to an integrated luminosity of $3.0\invnb$, while the fill recorded with the magnetic field pointing downwards comprises ${134 \times 10^6}$ events and corresponds to an integrated luminosity of $2.4\invnb$. The average numbers of collisions in a bunch crossing are $0.9$ and $0.7$ for these two fills, respectively. The data analysis is independent of the number of collisions per event, which means that events with multiple collisions are not treated differently. The results are obtained from the combined data sample, but as a cross-check, the analysis is also performed separately for each fill. To measure the background from interactions of the beams with residual gas in the beam pipe, beam-gas collisions are used, where only one of the two beams traverses the detector. Such collisions were also collected for each fill.

Simulation is required to model the effects of the imposed selection requirements and to study the background contributions. In the simulation, $pp$ collisions are generated using \pythia~8.1~\cite{Sjostrand:2014zea,*Sjostrand:2006za} with a specific \lhcb configuration~\cite{LHCb-PROC-2010-056}. Decays of unstable particles are described by \evtgen~\cite{Lange:2001uf}, in which final-state radiation is generated using \photos~\cite{davidson2015photos}. The interaction of the generated particles with the detector, and its response, are implemented using the \geant toolkit~\cite{Allison:2006ve, *Agostinelli:2002hh} as described in Ref.~\cite{LHCb-PROC-2011-006}. For each of the two magnetic-field configurations, a trigger-unbiased sample containing $10^7$ events was simulated.

\section{Analysis strategy}
\label{sec:analysis_strategy}

The differential cross-section is computed as
\begin{equation}
  \frac{\deriv^2 \sigma}{\deriv \eta \, \deriv \pt} \equiv \frac{n}{\lum \, \Delta \eta \, \Delta \pt} \,
  \label{eqn:differential_cross_section}
\end{equation}
where $n$ is the number of prompt long-lived charged particles produced in beam-beam collisions in a pseudorapidity interval with a width of $\Delta \eta$ and a transverse-momentum interval with a width of $\Delta \pt$ that is obtained from a data set corresponding to an integrated luminosity of \lum. It is computed in finite intervals of $\eta$ and \pt. The count $n$ cannot be obtained directly, it is computed from the observable number of recorded tracks after applying corrections, as further detailed below. There is no special treatment for events with multiple collisions per bunch crossing in this approach; every collision adds to the particle count and to the luminosity.

Tracks that traverse the entire tracking system are selected as candidates for prompt long-lived charged particles. Among these tracks, some are fakes, which do not correspond to a real particle. Fake tracks mostly originate from random matches of unrelated track segments upstream and downstream of the magnet. Two kinds of fake tracks are distinguished, depending on whether the fake track forms in isolation or nearby a real track. The first kind is reduced by imposing a requirement on the fake-track probability, $P_\mathrm{fake}$, provided by a neural-network-based algorithm~\cite{DeCian:2255039}, but the remaining contribution is still non-negligible. The second kind is always accompanied by a nearby track and through that signature already detected. It is suppressed by the track-reconstruction software below the level of $0.1\,\%$ in all kinematic intervals and therefore negligible in this analysis. Non-prompt tracks passing the selection are another source of background. These tracks originate from interactions of particles with the detector material or from decays of long-lived particles. Interactions of the beams with residual gas are another source of background. These background contributions are further discussed in Sect.~\ref{sec:background_contributions} as well as how much they contribute to the number of candidates.

Consequently, the number of candidate tracks, $n_\mathrm{cand}$, is related to the number of signal particles, $n$, according to
\begin{equation}
  n_\mathrm{cand} = \eps \, n + \sum_i n_i \, ,
\end{equation}
where \eps denotes the total efficiency, \ie the product of the geometric acceptance of the detector, the track-reconstruction efficiency and the selection efficiency, and the sum includes the numbers of background tracks, $n_i$, from source $i$. The values of \eps and $n_i$ are taken from simulation, except for the background from beam-gas interactions, $n_\mathrm{gas}$, which is determined from a control sample of pure beam-gas events. Control measurements are performed to correct for imperfect modelling in the simulation. For this purpose, observables, $\mathcal{P}_i$, that are proportional to $n_i$ are chosen as proxies. The background contributions and the corresponding proxies are described in Sect.~\ref{sec:background_contributions}. The ratio of the background counts in data and simulation is assumed to be equal to the ratio of the proxies in data and simulation,
\begin{equation}
  \frac{n_i}{n_{i,\,\mathrm{sim}}} = \frac{\mathcal{P}_i}{\mathcal{P}_{i,\,\mathrm{sim}}} \equiv R_i \, ,
\end{equation}
which allows the background count in data to be estimated as ${n_i = R_i \, n_{i,\,\mathrm{sim}}}$.

In summary, the number of signal particles in Eq.~\eqref{eqn:differential_cross_section} can be expressed as
\begin{equation}
  n = \frac{1}{R_\eps \, \eps_\mathrm{sim}} \left(n_\mathrm{cand} - \sum_i R_i \, n_{i,\,\mathrm{sim}} - n_\mathrm{gas}\right) ,
  \label{eqn:particle_count}
\end{equation}
where $R_\eps$ is a correction to the total efficiency in simulation.

Due to the slow variation of the differential cross-section in $\eta$, only six intervals in $\eta$ are used with ${\Delta \eta = 0.5}$. The width of the last interval, ${\eta \in [4.5, 4.8)}$, is reduced to match the acceptance of the tracking system. Since the \pt spectrum has a power-law shape, 50 logarithmic intervals are used in the range ${\pt \in [10, 10\,000)\mevc}$ with ${\Delta \!\log_{10}(\pt / (\!\mevc)) = 0.06}$. The lower edge of this grid extends beyond the fiducial range of \pt intervals that are used in the analysis, which starts at $80 \mevc$ or higher, depending on the $\eta$ interval. The lower limit is determined by the fact that a particle needs approximately $2\gevc$ to reach the tracking stations downstream of the magnet. In this analysis, an even tighter requirement ${\ptot > 5\gevc}$ is used, since a correction of the efficiency, which is described in Sect.~\ref{sec:efficiencies}, is based on the two control measurements presented in Refs.~\cite{LHCb-DP-2013-002,LHCb-PAPER-2011-037}, which both applied this tighter requirement.

The track-reconstruction efficiency depends on the detector occupancy. To ensure that the simulation reproduces the occupancy observed in data, weights are assigned to the simulated events. The number of tracks that traverse the entire tracking system is used as a proxy for the occupancy. Simulated events are weighted by the ratio of the distributions of the number of these tracks per event in data and simulation. As a positive side effect, this also adjusts the normalisation of the simulated sample with respect to the data sample. Whenever the simulation is mentioned in the following, it is always the occupancy-weighted simulation.

The two simulated samples are each weighted to reproduce the occupancy of the \lhc fill with the same magnetic-field configuration. As a cross-check, the chosen proxy for the occupancy is compared to an alternative proxy, the number of hits in the scintillating-pad detector, which is not affected by possible artefacts of the track reconstruction. A linear relation is observed between the number of tracks that traverse the entire tracking system and the number of hits in the scintillating-pad detector, confirming this choice of the proxy for the detector occupancy.

The detector resolution distorts the measured kinematic distributions, by inducing migration between different $(\eta, \pt)$ intervals. This distortion is between $1\,\%$ and $2\,\%$ in the simulation, depending on the kinematic interval. The simulated efficiency, $\eps_\mathrm{sim}$, implicitly takes these migration effects into account. It is calculated from simulation as the ratio of the number of candidate tracks that can be associated with signal particles and the respective number of signal particles, where the candidates are sorted into intervals using the reconstructed momenta, while the signal particles are sorted into intervals using the generated momenta. Deviations between the shapes of the kinematic distributions of charged particles in data and simulation can introduce a bias in this approach, but the bias in this case is estimated to be at the order of $0.01\,\%$ and therefore negligible.

The \textsc{ROOT}~\cite{Brun:1997pa} and \lhcb~\cite{Corti:2006yx,Tsaregorodtsev:2010zz} software frameworks are used for the initial data preparation, while the analysis is written in the \textsc{Python} language with standard scientific packages~\cite{Hunter:2007ouj,Lam:2020a,Harris:2020xlr,Virtanen:2019joe,Koester:2012a,Meurer:2017yhf} and high-energy-physics-specific packages~\cite{Schreiner:2021a,Dembinski:2020a,Rodrigues:2020a,Pivarski:2020a} from the \textsc{Scikit-HEP} project~\cite{Rodrigues:2020syo}.

\section{Efficiencies}
\label{sec:efficiencies}

\begin{figure}
  \centering
  \includegraphics[width=\textwidth]{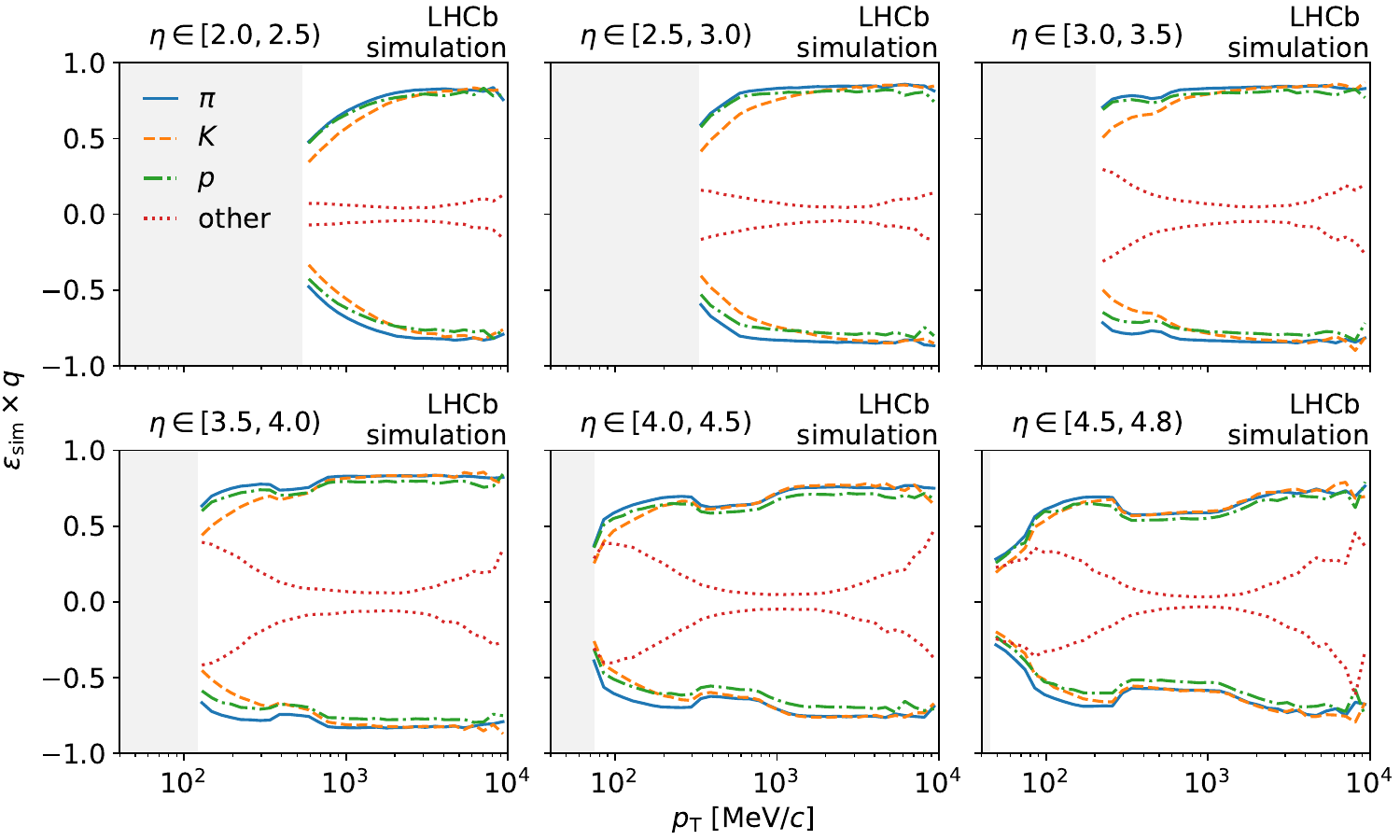}
  \caption{Efficiencies for different particle species in intervals of $\eta$ and as a function of \pt for the simulated sample generated with the magnetic field pointing upwards. The product of the efficiency and the particle charge in units of the elementary charge is shown to separate values for oppositely charged particles. The light-grey areas indicate the limit of the kinematic acceptance.}
  \label{fig:efficiency_per_particle_simple_4201}
\end{figure}

The total efficiency to observe prompt long-lived charged particles depends on the geometric acceptance of the detector, the track-reconstruction efficiency, and the particle loss due to decays and interactions with the detector material. The losses depend on the particle composition and the amount of traversed material. The composition of the prompt long-lived charged particles has an impact on the total efficiency, because pions and protons have different interaction lengths, for example.

In this study, the composition is broken into four particle classes: charged pions, charged kaons, protons, and other long-lived charged particles (\en, \mun, \Sigmap, \Sigmam, \Xim and \Omegam). The detector acceptance and track-reconstruction efficiency are taken from simulation, but corrections from control measurements are applied, which are detailed below. The simulated efficiencies, $\eps_\mathrm{sim}$, for the four particle classes are shown in Fig.~\ref{fig:efficiency_per_particle_simple_4201}. The efficiency for the last class is generally lower, because \Sigmap, \Sigmam, \Xim and \Omegam baryons and their antiparticles have the shortest lifetimes. At low \pt, the efficiency is higher, since the relative yield of these hadrons is small and leptons contribute more, which have high tracking efficiencies. At high \pt, the efficiency increases again, because the lifetimes of the strange hadrons are Lorentz boosted. This particle class contributes little to the total efficiency, since the particles in the class are less abundantly produced compared to the particles of the other three classes. A dip in the efficiency can be seen in the ${\eta \in [3.0, 3.5)}$ interval at ${\pt \approx 500\mevc}$, which becomes more pronounced in the following $\eta$ intervals. This feature is due to an increased amount of material that particles with certain trajectories have to pass~\cite{Needham:2007iz}. Efficiencies for particles and antiparticles are slightly different due to the different hadronic cross-sections. The simulated efficiency is validated against data. Small corrections for differences in the tracking efficiency and for differences in the hadron composition are applied in this analysis.

The track-reconstruction efficiency is corrected based on results of a separate control measurement~\cite{LHCb-DP-2013-002}, in which muon tracks from \decay{\jpsi}{\mumu} decays were studied to determine ratios of the track-reconstruction efficiencies in data and simulation in intervals of $\eta$ and \ptot. A linear transformation matrix is built to map the results of the control measurement to the $\eta$ and \pt intervals used in this analysis. Non-uniform track density is accounted for in this transformation. The obtained efficiency ratios are identical for both particle charges and are used as the first component of the correction to the efficiency in simulation. The uncertainty of this component is between $1\,\%$ and $5\,\%$ over the kinematic range covered by this analysis.

The simulated particle composition is then corrected by adjusting the relative yield of each particle class to match data. Since the composition has not yet been measured in \proton\proton collisions at ${\sqs = 13\tev}$, \lhcb measurements of ratios of prompt hadron production are extrapolated from \proton\proton collisions at ${\sqs = 0.9\tev}$ and $7\tev$~\cite{LHCb-PAPER-2011-037}. The hadron ratios $\antiproton / \proton$, ${(\Kp + \Km) / (\pip + \pim)}$ and ${(\proton + \antiproton) / (\pip + \pim)}$ depend on $\eta$ and \pt, and are separately extrapolated to $13\tev$ in each $\eta$ and \pt interval with a linear function in $\ln \sqs$. The largest deviations of up to $40\,\%$ are observed between the extrapolated and simulated ${(\proton + \antiproton) / (\pip + \pim)}$ ratios.

The extrapolated ratios obtained from the first step cover only three \pt intervals and do not cover the intervals ${\eta \in [2.0, 2.5)}$ and ${\eta \in [4.5, 4.8)}$. To mitigate this, double ratios are computed from the extrapolated hadron ratios and the corresponding hadron ratios in this simulation for each $\eta$ and \pt interval. Forming double ratios removes most of the $\eta$ and \pt dependence of the hadron ratios, which makes their extrapolation more robust. The double ratios for the intervals ${\eta \in [2.0, 2.5)}$ and ${\eta \in [4.5, 4.8)}$ are taken from the corresponding adjacent $\eta$ intervals for further analysis.

In a final step, correction factors for the counts of long-lived particles in the four classes are introduced. These correction factors are parameterised as linear functions of $\ln \pt$ separately for each $\eta$ interval, particle class, and charge. The parameters of these correction functions are optimised in a least-squares fit so that the double ratios formed from the corrected and the original hadron ratios in this simulation approach the double ratios obtained in the previous step, under the constraint that the total number of particles per charge remains the same. Since this optimisation problem is underconstrained by the extrapolated ratios, Gaussian penalty terms are introduced as a regularisation to suppress deviations above $5\,\%$ of the corrected counts from their initial values. The composition-corrected efficiency in simulation is then computed by summing the products of the efficiencies for the four particle classes with their corrected fractions. Since this correction involves considerable modelling, the systematic uncertainty dominates over the statistical uncertainty. Half of the correction is assigned as a systematic uncertainty as a conservative estimate, which is at most $2.5\,\%$.

\begin{figure}
  \centering
  \includegraphics[width=\textwidth]{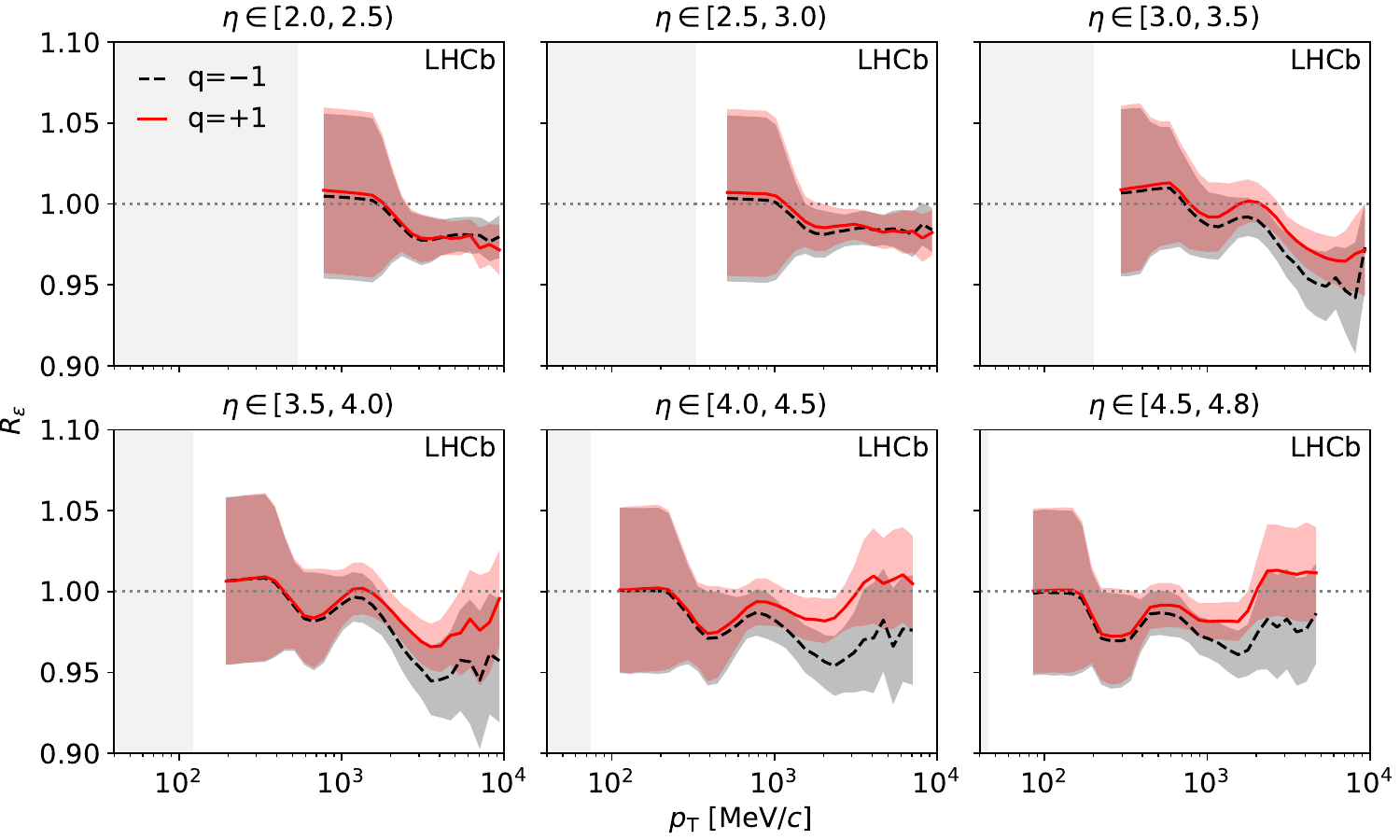}
  \caption{Efficiency corrections, $R_\eps$, for positively and negatively charged particles in intervals of $\eta$ and as a function of \pt for the simulated sample generated with the magnetic field pointing upwards. The bands indicate the systematic uncertainty. The light-grey areas represent the limit of the kinematic acceptance. The gap between this limit and the correction is due to the tighter requirement of ${\ptot > 5\gevc}$ applied to the control measurements on which this correction is based.}
  \label{fig:efficiency_ratio_4201}
\end{figure}

The combined correction of the effects discussed so far for the simulated efficiency is shown in Fig.~\ref{fig:efficiency_ratio_4201} and reduces the original efficiency in simulation by up to $5\,\%$ for negatively charged particles. This reduction arises primarily from a relative increase in the fourth particle class of the particle composition, which has a lower efficiency. The step-like structures arise from the comparably coarse $(\eta, p)$ intervals over which the control measurement for the track-reconstruction efficiency is performed. The transfer to an $(\eta, \pt)$ grid smoothens these steps, but they remain visible. Figure~\ref{fig:efficiency_ratio_4201} also defines the usable \pt range in each $\eta$ interval. The \pt intervals without an efficiency correction are not used to compute final results.

Material interactions contribute to the systematic uncertainty of the corrected efficiency as the simulated amount of material has an uncertainty of $10\,\%$, as described in Ref.~\cite{LHCb-DP-2013-002}. The hadronic losses are proportional to the amount of traversed material and therefore carry the same relative uncertainty. The interaction losses of charged pions and charged kaons are estimated to be $14\,\%$ and $11\,\%$ in Ref.~\cite{LHCb-DP-2013-002}, respectively. For protons, the loss is between $20\,\%$ and $30\,\%$ across the full kinematic range. This loss is estimated from the difference between the efficiency for protons in simulation and that for muons. Since protons are stable and muon decays inside \lhcb are negligible, any additional loss of protons relative to muons is caused by material interactions. Hadronic loss for the species in the fourth class of long-lived particles is negligible, as they are either leptons or their loss is dominated by decays. Overall, the hadronic losses have uncertainties of up to $3\,\%$, depending on the particle species. Yet, the final uncertainty on the efficiency due to material interactions amounts to only $1\,\%$, because the uncertainties contribute proportional to the abundance of the corresponding particle class.

The total systematic uncertainty of the corrected efficiency is the sum of the contributions from the correction of the track-reconstruction efficiency, the particle-composition correction, material interactions and the statistical uncertainty of the simulated samples. The latter ranges between $0.06\,\%$ and $2\,\%$, depending on the kinematic interval, while the total uncertainty ranges from $0.9\,\%$ to $5.1\,\%$.

\section{Background contributions}
\label{sec:background_contributions}

The background induced by beam-gas interactions is determined from the number of candidate tracks produced in a control sample of recorded pure beam-gas events. Both the configuration where the beam travels from the vertex detector towards the muon system and the opposite configuration are included. The contributions from these configurations are scaled to the corresponding number of recorded beam-beam events. The fraction of candidate tracks originating from beam-gas interactions is found to be $1\,\%$ for the \lhc fill recorded with the magnetic field pointing downwards, independent of the kinematic interval. For the fill recorded with the magnetic field pointing upwards, the fraction is below the level of $0.1\,\%$.

\begin{figure}
  \centering
  \includegraphics[width=\textwidth]{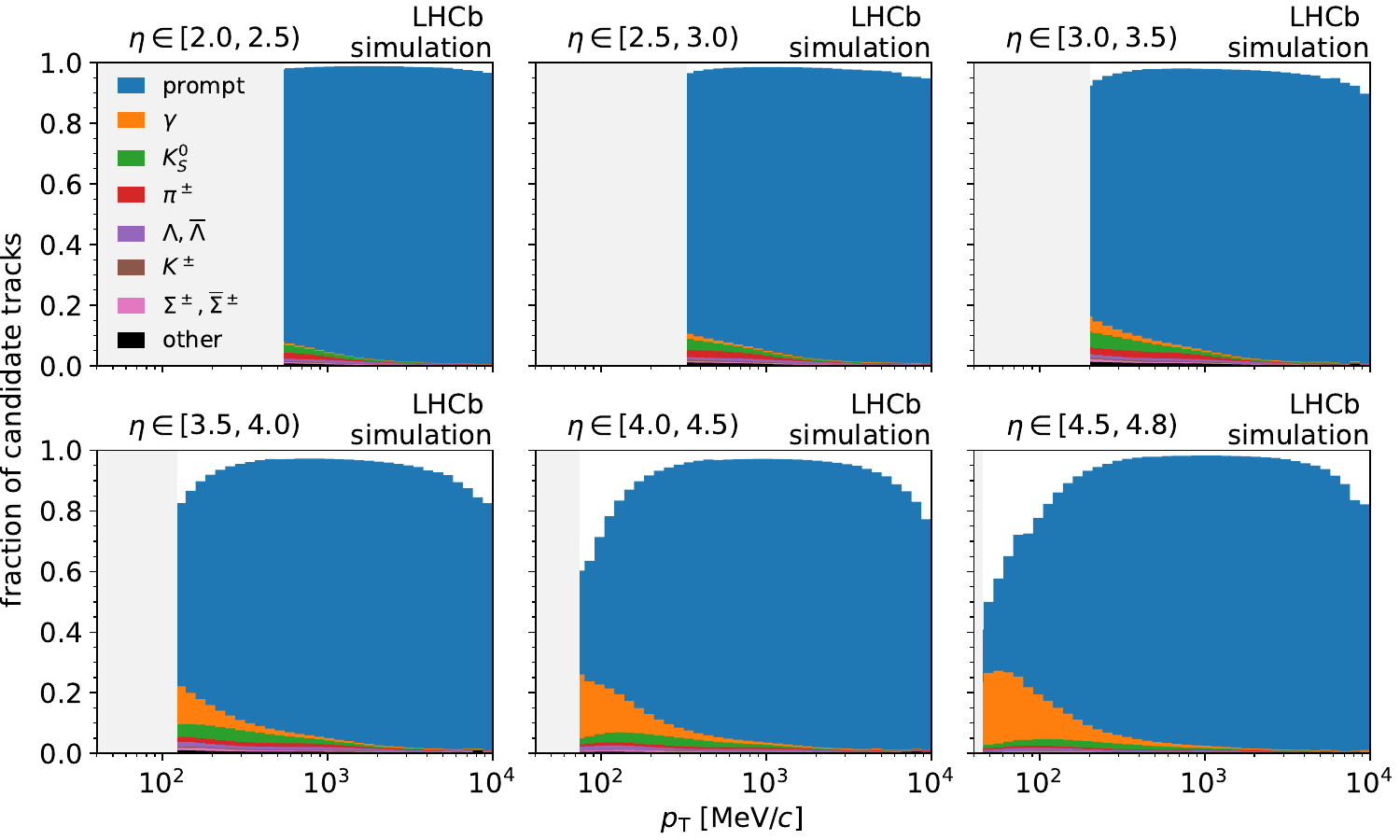}
  \caption{Origins of the candidate tracks in intervals of $\eta$ and as a function of \pt for the simulated sample generated with the magnetic field pointing upwards. The \textit{prompt} category refers to the signal tracks, while the other categories correspond to non-prompt tracks originating from the listed parent particles and include both decays and material interactions. Fake tracks are indicated by the white areas above the stacked histograms. The light-grey areas represent the limit of the kinematic acceptance.}
  \label{fig:background_origin_candidate_sim09b_up}
\end{figure}

The origins of the candidate tracks in simulation are shown in Fig.~\ref{fig:background_origin_candidate_sim09b_up}. Prompt long-lived charged particles constitute more than $85\,\%$ of the candidate tracks around ${\pt = 1\gevc}$. Towards lower or higher values of \pt, the background contributions increase. At high \pt, fake tracks are the largest background, while fake tracks and tracks from photon conversions dominate at low \pt. These background contributions are quantified using simulation and adjusted with correction factors from proxies obtained from data and simulation. The background contributions that are sufficiently large to require a correction are: fake tracks, tracks originating from material interactions of charged pions and from photon conversions, and tracks produced in decays of strange hadrons. In the following subsections, the proxies constructed for these sources of background are presented. The remaining sources, which are not adjusted, are combined and a conservative uncertainty of $50\,\%$ is assigned for their contribution. The uncertainty of the differential cross-section resulting from this assumption is negligible, as the contribution from the remaining sources of background is small.

\subsection{Fake tracks}

The largest background at the edges of the analysed \pt range is due to isolated fake tracks, while the contribution of fake tracks accompanied by a nearby track is negligible, as discussed in Sect.\ \ref{sec:analysis_strategy}. These are characterised by having a low quality of the track fit and missing hits in instrumented detector parts. This information is combined with further variables from the tracking system to estimate a fake-track probability, $P_\mathrm{fake}$. The imposed selection requirement, ${P_\mathrm{fake} < 0.3}$, rejects between $40\,\%$ and $80\,\%$ of the fake tracks and retains $99\,\%$ of the real tracks. The remaining background from fake tracks, shown in Fig.~\ref{fig:background_origin_candidate_sim09b_up} as white areas above the stacked histograms, ranges between $1\,\%$ and $27\,\%$ in the effective \pt range derived from Fig.~\ref{fig:efficiency_ratio_4201}.

\begin{figure}
  \centering
  \includegraphics[width=0.49\textwidth]{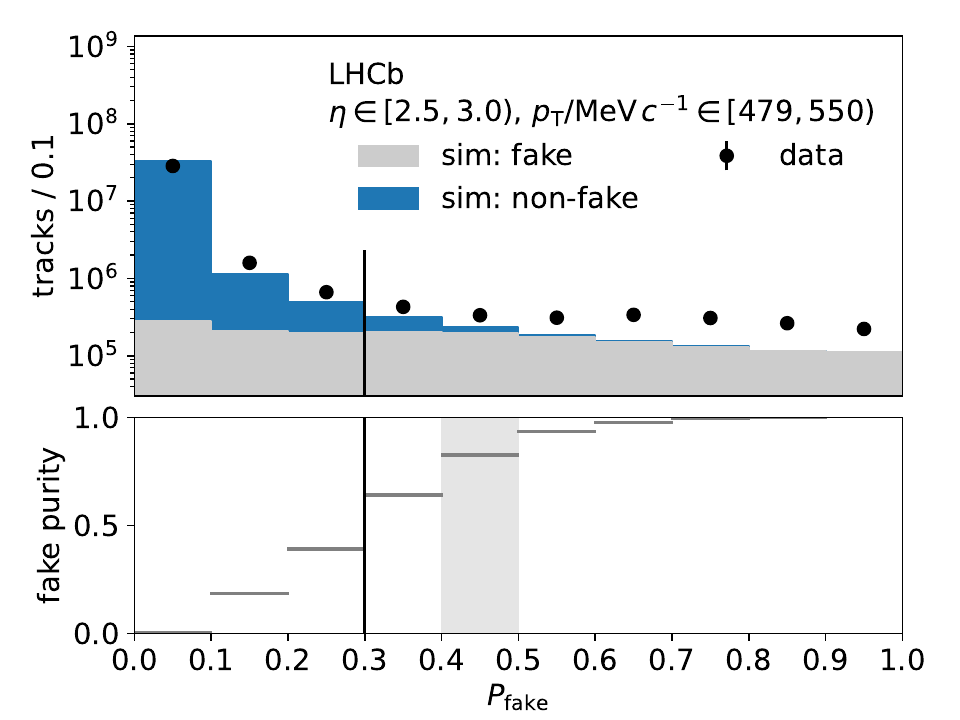}
  \includegraphics[width=0.49\textwidth]{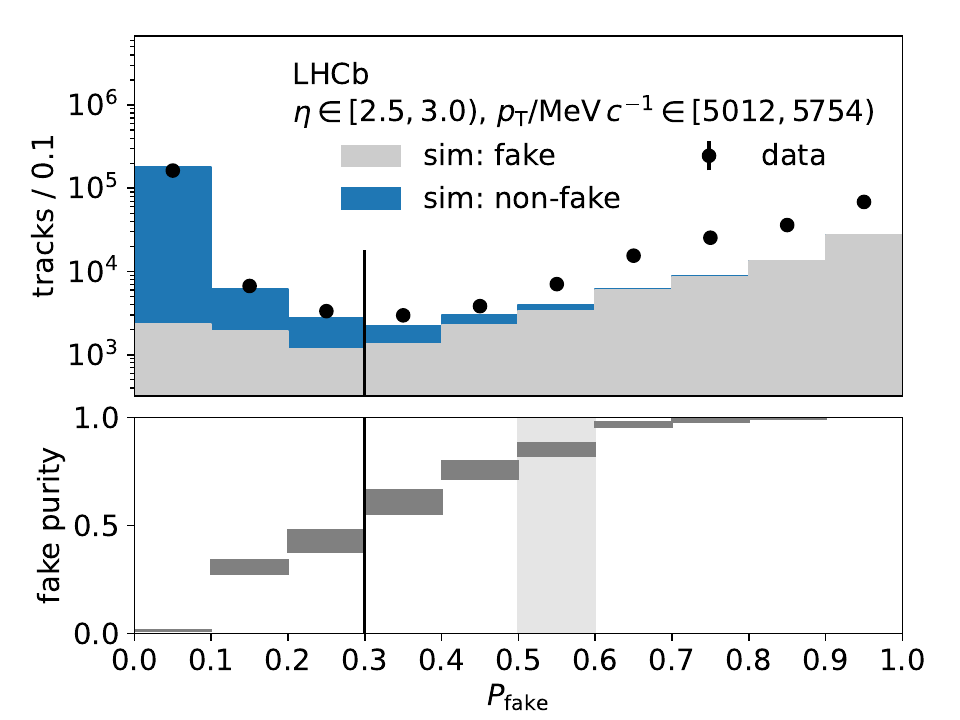}
  \caption{Distributions of $P_\mathrm{fake}$ in a kinematic interval (left) at low \pt and (right) at high \pt for the \lhc fill recorded with the magnetic field pointing upwards and the simulated sample with the same magnetic-field configuration. The statistical uncertainty of the data is indicated by error bars, but they are not visible on these scales. The purity of the proxy is shown in the lower panels, where light-grey boxes represent the $P_\mathrm{fake}$ interval selected to determine the proxy. Vertical lines indicate the threshold ${P_\mathrm{fake} < 0.3}$ for candidate tracks. The normalisation of the simulated sample is described in Sect.~\ref{sec:analysis_strategy}.}
  \label{fig:1_28_4201}
\end{figure}

For each kinematic interval, data and simulation are divided into ten equal-width intervals in the $P_\mathrm{fake}$ variable. Tracks of both charges are combined in this analysis, since the background from fake tracks is charge symmetric. Distributions in two different kinematic intervals are given as examples in Fig.~\ref{fig:1_28_4201}. In simulation, a pure sample of fake tracks is present at high $P_\mathrm{fake}$ values. Generally, good agreement is found in the shapes of the $P_\mathrm{fake}$ distributions between data and simulation, but in simulation, fewer fake tracks are observed. This can be seen in Fig.~\ref{fig:1_28_4201}, where the intervals with ${P_\mathrm{fake} > 0.5}$ contain more entries in data compared to simulation.

Consequently, a correction is used to adjust the simulated contribution from fake tracks. The number of tracks in the first interval of the $P_\mathrm{fake}$ distribution above $0.3$, where the fake-track purity is greater than $80\,\%$, is chosen as a proxy for fake tracks. This choice balances two sources of systematic uncertainty. A $P_\mathrm{fake}$ interval with a high fake-track purity is required in order to select a proxy that is insensitive to possible differences in the number of signal tracks between data and simulation, which would favour the interval with the highest purity. However, selecting an interval close to the $P_\mathrm{fake}$ region of interest, ${P_\mathrm{fake} < 0.3}$, minimises the impact of shape differences in the $P_\mathrm{fake}$ distributions between data and simulation.

The uncertainty of the fake-track proxy is the quadratic sum of statistical and systematic contributions. The statistical uncertainty is propagated from the track counts in data and simulation. The systematic uncertainty is estimated by computing an alternative proxy. For this, the integral of the track counts in the selected $P_\mathrm{fake}$ interval and all the intervals above is used. The alternative proxy is more affected by shape differences and is therefore an upper limit on the systematic error. This type of systematic variation is modelled with a uniform distribution of deviations of which the upper limit corresponds to the alternative proxy, while the centre is the value of the default proxy. The uncertainty of the proxy is then given by the standard deviation of this distribution.

\begin{figure}
  \centering
  \includegraphics[width=\textwidth]{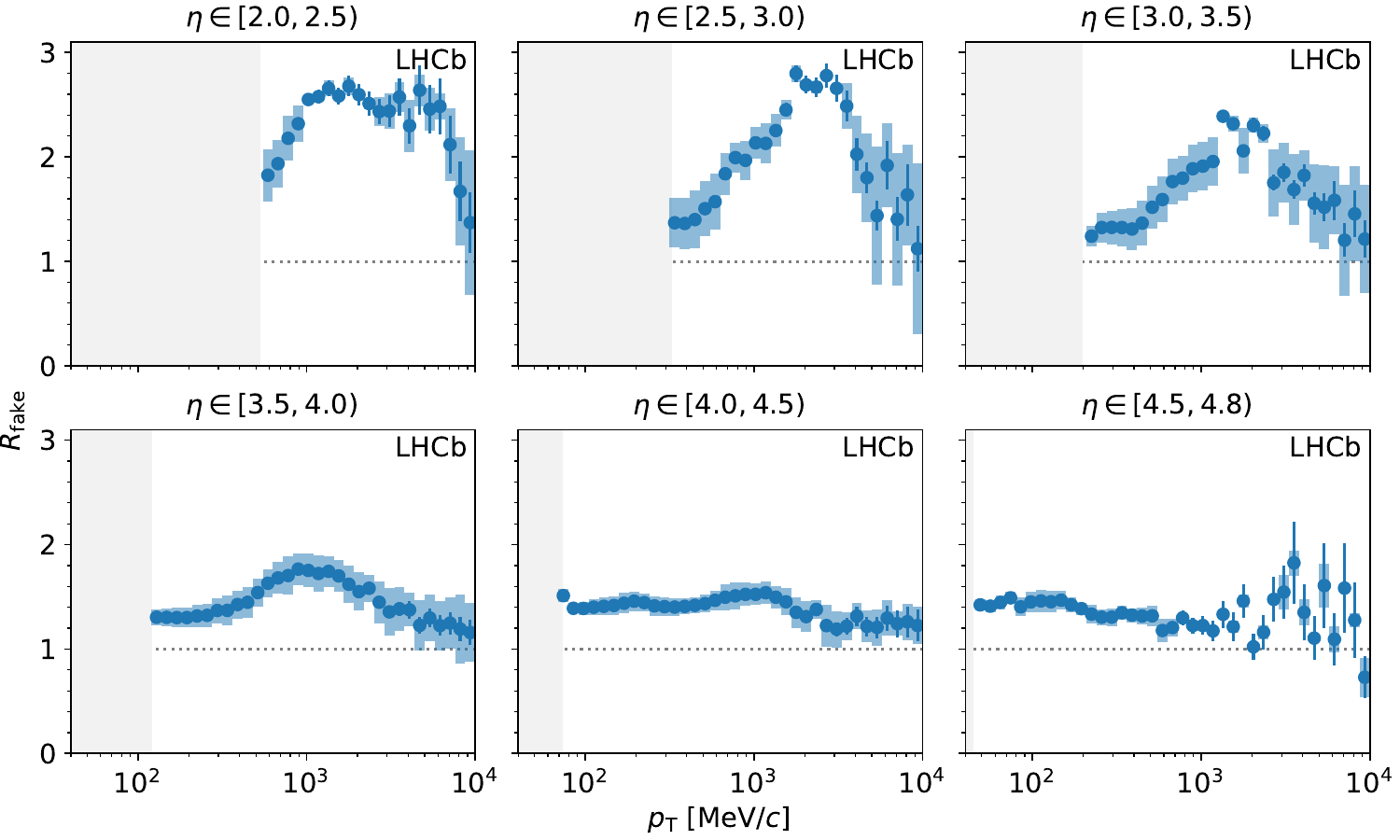}
  \caption{Ratio of the proxies for fake tracks in data and simulation in intervals of $\eta$ and as a function of \pt for the \lhc fill recorded with the magnetic field pointing upwards and the simulated sample with the same magnetic-field configuration. The error bars indicate the statistical uncertainty, and the boxes indicate the systematic uncertainty. The light-grey areas represent the limit of the kinematic acceptance.}
  \label{fig:fake_ratio_4201}
\end{figure}

The resulting proxy ratio is shown in Fig.~\ref{fig:fake_ratio_4201}. At low $\eta$, values of the proxy ratio of up to $2.8$ are observed, but the \pt intervals in which the proxy ratio deviates strongly from unity are also those where the background from fake tracks is very small, as it can be seen by comparing with Fig.~\ref{fig:background_origin_candidate_sim09b_up}. Where the background from fake tracks is large, the proxy ratio is between $1.2$ and $1.4$. In general, the ratio is smooth as a function of \pt. Discontinuities in the range ${\eta \in [2.5, 3.5)}$ are caused by changes of the $P_\mathrm{fake}$ interval chosen to determine the proxy.

\subsection{Material interactions}

Non-prompt tracks produced in material interactions constitute the second-largest source of background. As only tracks that traverse the entire tracking system are used, these are interactions occurring inside the vertex detector. Electrons originating from photon conversions, which populate mainly the kinematic intervals at high $\eta$ and low \pt, contribute up to $20\,\%$ to the candidate tracks, while charged particles stemming from hadronic material interactions of charged pions contribute less than $5\,\%$.

The number of tracks originating from interactions in the material is proportional to the product of the total particle flux, consisting primarily of charged pions, and the amount of traversed material. The number of tracks that originate from candidate vertices, in which three tracks converge and which pass certain selection criteria that are detailed below, is chosen as a proxy for this product. Vertices with three tracks are used because hadronic interactions frequently produce three or more charged particles, while decays into three or more charged particles are rare in the analysed data sample. This approach is similar to that presented in Ref.~\cite{LHCb-DP-2018-002}. Instead of defining a separate proxy for photon conversions, their simulated yield is scaled with the proxy for charged-pion interactions in the material. This is motivated by the observation that most photons originate from decays of neutral pions, which are produced in a fixed ratio to charged pions due to isospin symmetry. Furthermore, the conversion probability of photons and the interaction probability of charged pions are both proportional to the amount of material.

In each event, all unique combinations of three candidate tracks are formed first. Their point of closest approach is computed with a vertex fit minimising the sum of the squared distances of the tracks. This point is the candidate vertex of an interaction. To reduce combinatorial background, where three tracks are randomly associated with a common vertex, a lower limit is imposed on the distance of the vertex from the $z$ axis, which points from the radial centre of the vertex detector along the beam axis towards the muon system and has its origin close to the average primary vertex. This discards the region in which no material is present.

The purity of the obtained sample of vertices is further increased by applying requirements that are optimised using simulation. First, a simultaneous optimisation of a lower limit on the $z$ coordinate of the vertex and an upper limit on the sum of the squared distances of the tracks to their associated vertex, $d$, is performed without subdividing the sample intervals in $\eta$ and \pt. The $z$ requirement suppresses combinatorial background by selecting vertices downstream of the region in which most of the primary vertices are reconstructed. The upper limit on $d$ rejects random combinations by requiring that the tracks are compatible with originating from a single point. Second, an optimisation of a lower limit on the mass of the three-track combination is performed separately for each kinematic interval. As the species of the produced particles are not known, a zero-mass hypothesis is assigned to each track. The mass requirement further suppresses the combinatorial background. For both optimisations, the figure of merit ${S / \sqrt{S + B}}$ is used, where $S$ denotes the number of tracks that represent the signal for this proxy, and $B$ is the number of background tracks. Wider \pt intervals are used in this study to reduce fluctuations in the proxy ratio to be determined, created by successively merging groups of three adjacent \pt intervals. The requirements chosen are those that maximise the value of the figure of merit. This results in a lower limit on $z$ of $290\mm$ and an upper limit on $d$ of $0.1\mma$. The lower limits on the mass depend on the interval and vary around 200\mevcc.

After the application of all optimised requirements, the contribution from charged pions is approximately equal in size to that from all other particles combined, \eg kaons or protons. For each kinematic interval and each particle charge, the number of tracks that are included in selected vertices is used as the proxy. In the first two $\eta$ intervals and around ${\pt = 500\mevc}$, the highest values of the purity of charged-pion interactions are obtained, which range from $40\,\%$ to $45\,\%$. Towards higher values of $\eta$ and lower or higher values of \pt, the purity decreases.

The proxy ratio is independent of the purity for kinematic intervals with purities above $30\,\%$. Therefore, the value of the proxy ratio, $R_\mathrm{mat}$, is only used if the purity in the kinematic interval under consideration is greater than the threshold value of $30\,\%$. Moreover, an upper limit of $30\,\%$ is imposed on the statistical uncertainty of the purity to reject intervals that contain only a small number of tracks. As a substitute for the values of the proxy ratio in the intervals below the purity threshold, the median of the values of the proxy ratio in the intervals above the threshold is computed. The uncertainty of this median value is estimated assuming a uniform distribution covering the observed range of $R_\mathrm{mat}$.

\begin{figure}
  \centering
  \includegraphics[width=\textwidth]{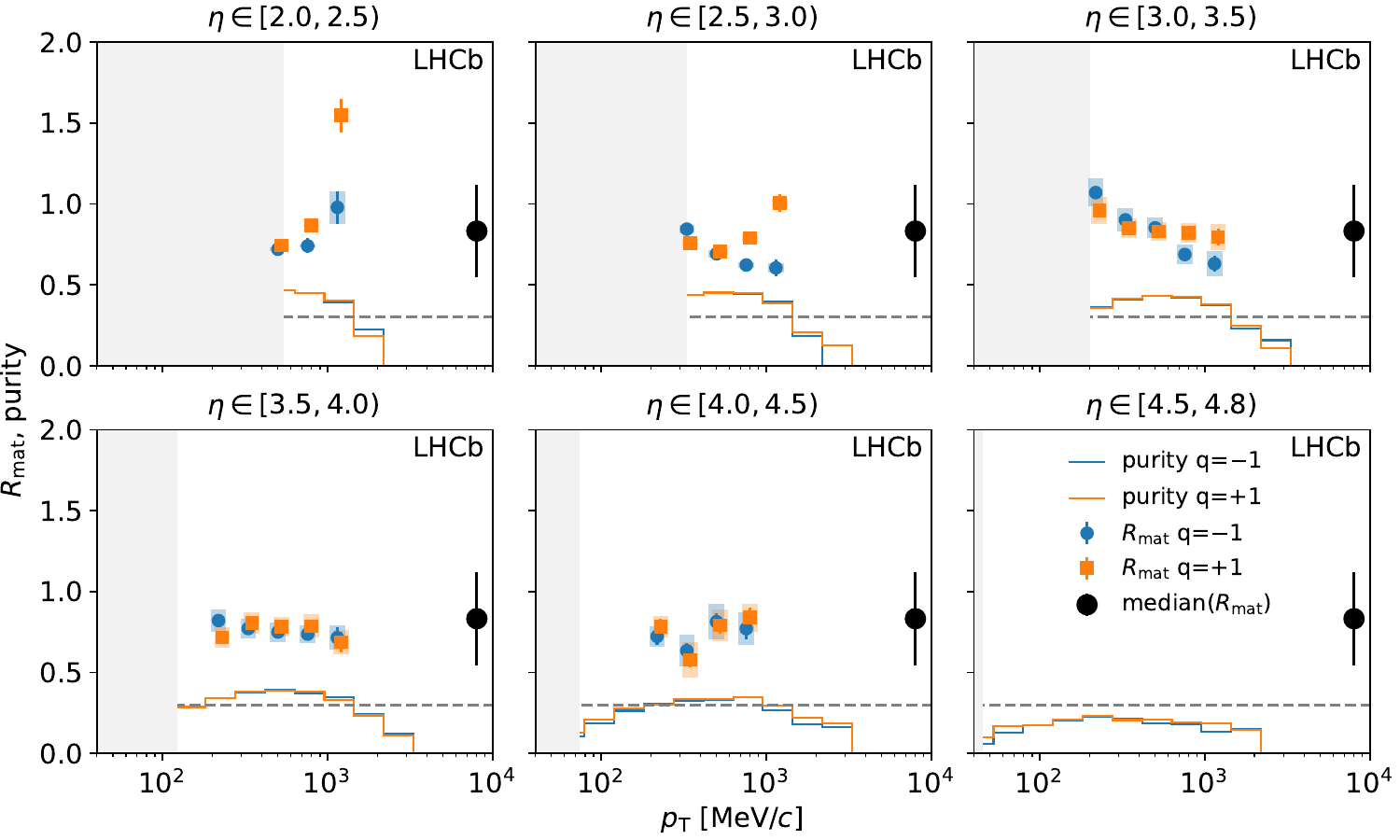}
  \caption{Ratio of the proxies for material interactions in data and simulation and purity of the proxy in intervals of $\eta$ and as a function of \pt for the \lhc fill recorded with the magnetic field pointing upwards and the simulated sample with the same magnetic-field configuration. The quantities are shown separately for positively and negatively charged particles. The error bars indicate the statistical uncertainty, and the boxes indicate the systematic uncertainty. The points are slightly displaced horizontally within the \pt intervals to increase the visibility. The dashed grey lines represent the purity threshold. The larger points indicate the median value computed from the intervals above the threshold. The light-grey areas represent the limit of the kinematic acceptance.}
  \label{fig:material_ratio_simple_4201}
\end{figure}

The systematic uncertainty of the proxy for material interactions is determined by loosening and tightening the optimised $z$ and $d$ requirements and computing alternative proxy ratios, $R_\mathrm{mat}^\prime$. The mass requirement is not varied, as a good agreement is found in the shapes of the distributions between data and simulation in all kinematic intervals. Deviations in some of the alternative proxies cannot be explained by statistical fluctuations and are therefore considered as a systematic uncertainty. The deviation of an alternative is considered significant if the condition
\begin{equation}
  (R_\mathrm{mat}^\prime - R_\mathrm{mat})^2 > \left|\sigma^2(R_\mathrm{mat}^\prime) - \sigma^2(R_\mathrm{mat})\right|
\end{equation}
is fulfilled, where $\sigma^2$ denotes the variance, which is a consequence of using the same data set to determine the default proxy ratio and the alternatives. Following Ref.~\cite{Barlow:2002yb}, the systematic uncertainty is taken as the standard deviation of the variations that differ significantly and the initial ratio is replaced with the mean of the significant variations. This procedure is applied to the values of the proxy ratio in the kinematic intervals above the purity threshold and to the median value computed for the other intervals. The values and the resulting uncertainties of the proxy ratio are shown in Fig.~\ref{fig:material_ratio_simple_4201}.

\subsection{Strange-hadron decays}

Up to $7\,\%$ of the candidate tracks are non-prompt tracks produced in decays of strange hadrons inside the vertex detector. Approximately $80\,\%$ of these tracks originate from \decay{\KS}{\pip\pim}, \decay{\Lz}{\proton\pim} and \decay{\Lbar}{\antiproton\pip} decays. The number of \KS mesons and \Lz and \Lbar baryons, of which both decay products create a candidate track, is chosen as a proxy for this background.

Candidate decays are obtained by forming pairs of oppositely charged candidate tracks. A small set of loose requirements is used to maintain high selection efficiency for strange-hadron decays. The distance of closest approach of the two tracks is required to be less than $1\mm$ and the distance along the $z$ axis of their point of closest approach from the average primary vertex is required to be greater than $150\mm$. For \KS candidates, the mass is computed by assigning the pion-mass hypothesis to both tracks, while for \Lz and \Lbar candidates, the proton- and pion-mass hypotheses are assigned to the tracks.

\begin{figure}
  \centering
  \includegraphics[width=0.49\textwidth]{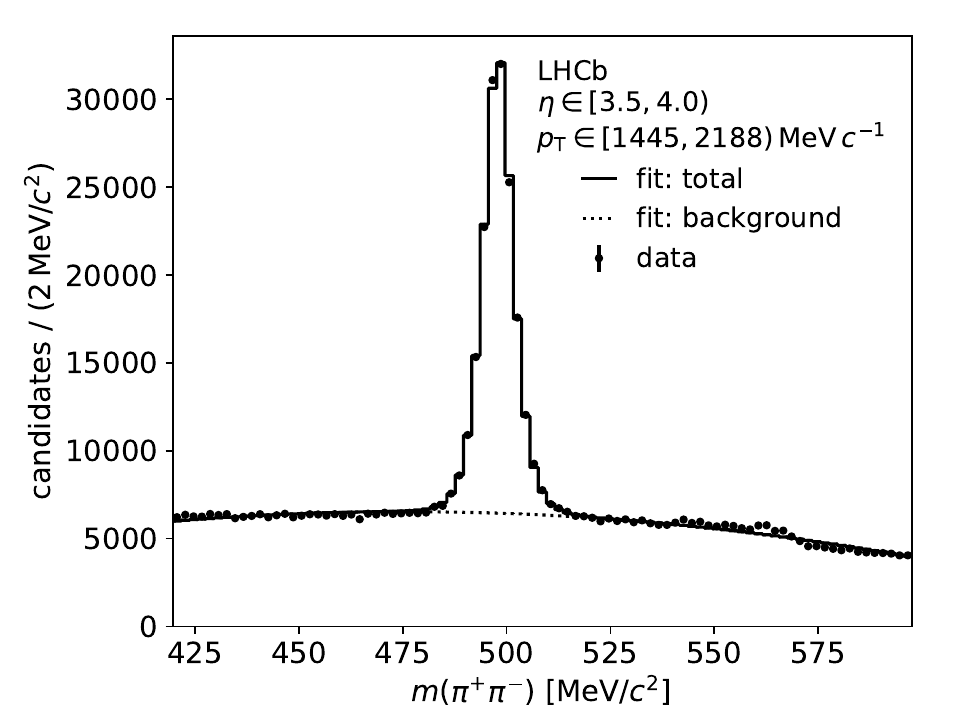}
  \includegraphics[width=0.49\textwidth]{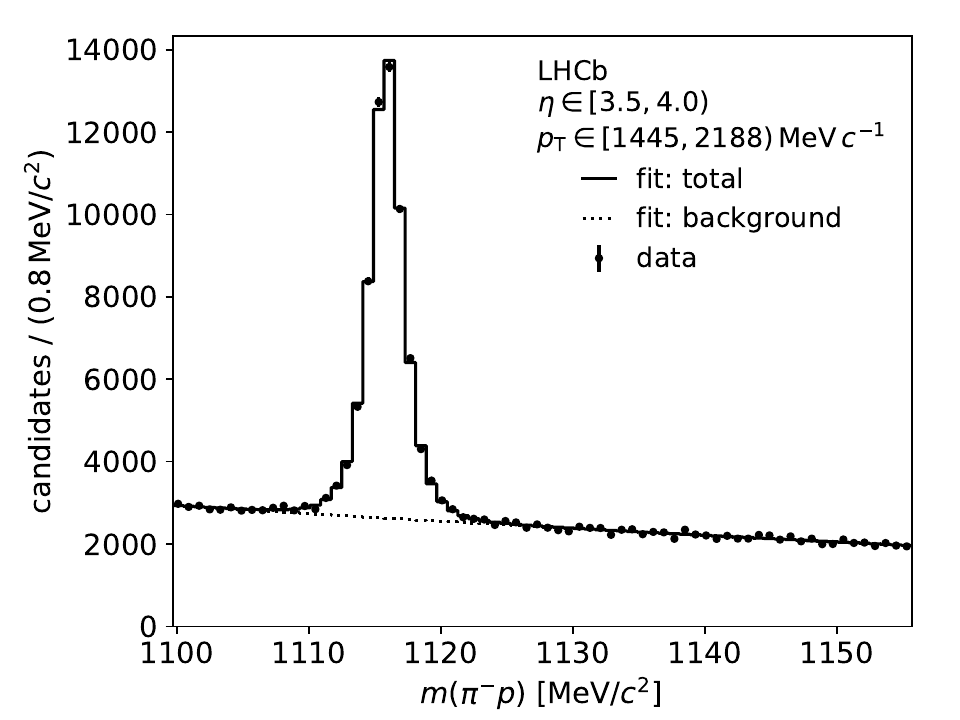}
  \includegraphics[width=0.49\textwidth]{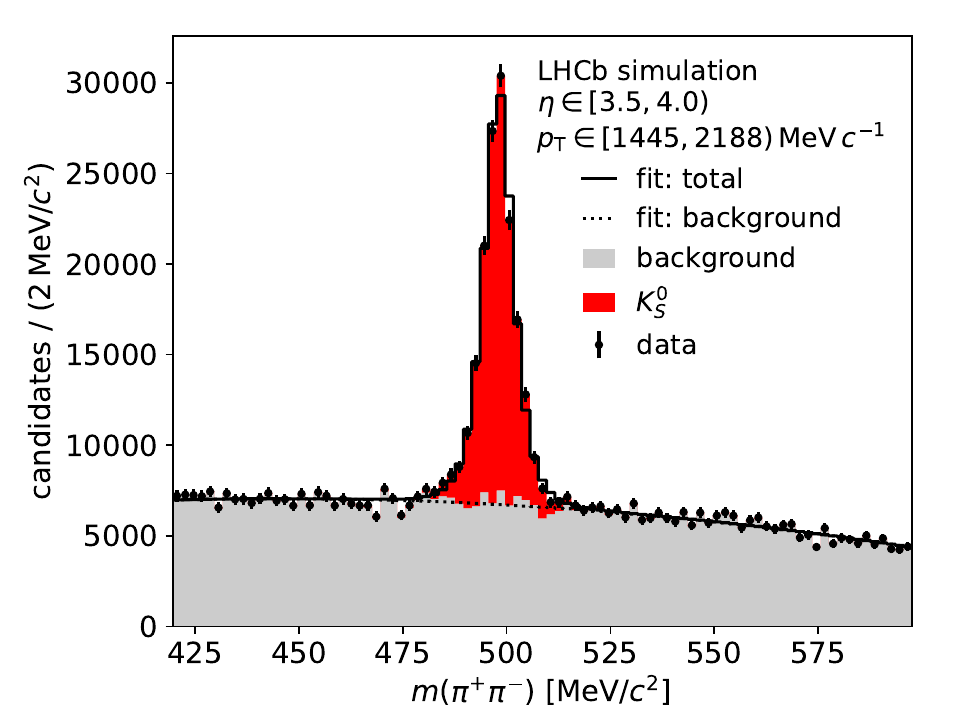}
  \includegraphics[width=0.49\textwidth]{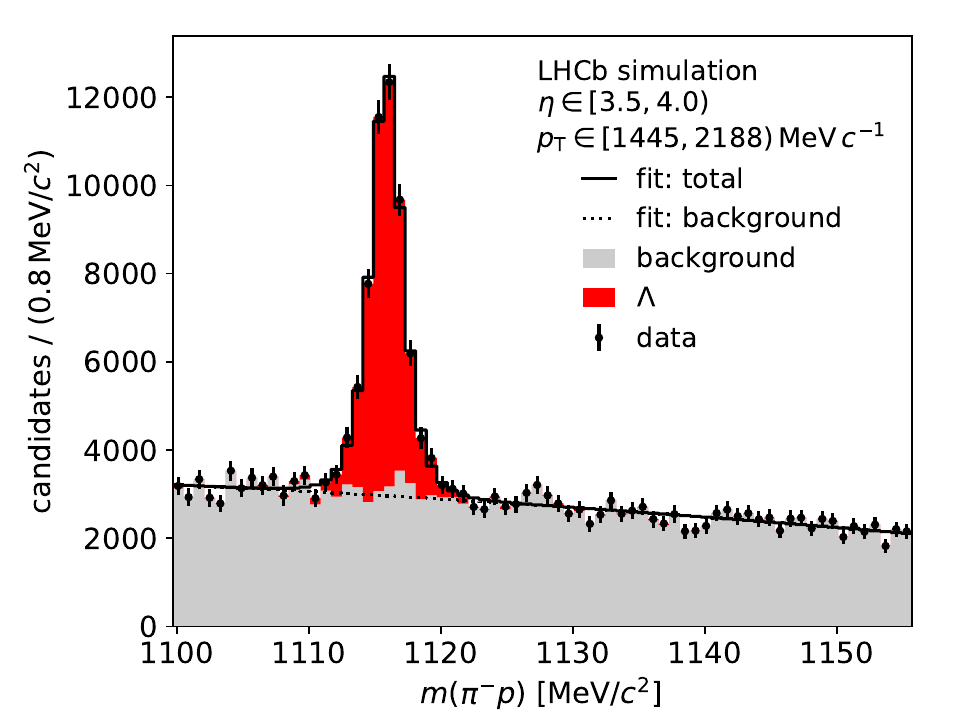}
  \caption{Mass distributions of (left) \KS and (right) \Lz candidates in one kinematic interval for (top) the \lhc fill recorded with the magnetic field pointing upwards and (bottom) the simulated sample with the same magnetic-field configuration. Solid lines indicate the total fit and dotted lines the background. In the case of simulation, the signal and background contributions are shown for illustration, but this information is not used in the analysis.}
  \label{fig:K_S_0_3_12_4201}
\end{figure}

The invariant-mass distributions of \KS and \Lz candidates in one kinematic interval are shown as examples in Fig.~\ref{fig:K_S_0_3_12_4201}. Extended-maximum-likelihood fits~\cite{Barlow:1990vc} to the mass distributions of the \KS, \Lz and \Lbar candidates in data and simulation are performed separately for each $(\eta, \pt)$ interval of their kinematic distributions. As in the case of the proxy for material interactions, wider \pt intervals are used in this study, obtained by merging groups of three adjacent \pt intervals. In the fits, the \KS, \Lz or \Lbar signal is modelled with a nonstandardised Student's t-distribution with free location and width parameters. The background, which is only combinatorial, is modelled with a third-degree Bernstein polynomial. The yields in each interval of the mass distributions are modelled with Poisson distributions for the unweighted histograms in data and with scaled Poisson distributions~\cite{Bohm:2013gla} for the occupancy-weighted histograms in simulation.

\begin{figure}
  \centering
  \includegraphics[width=\textwidth]{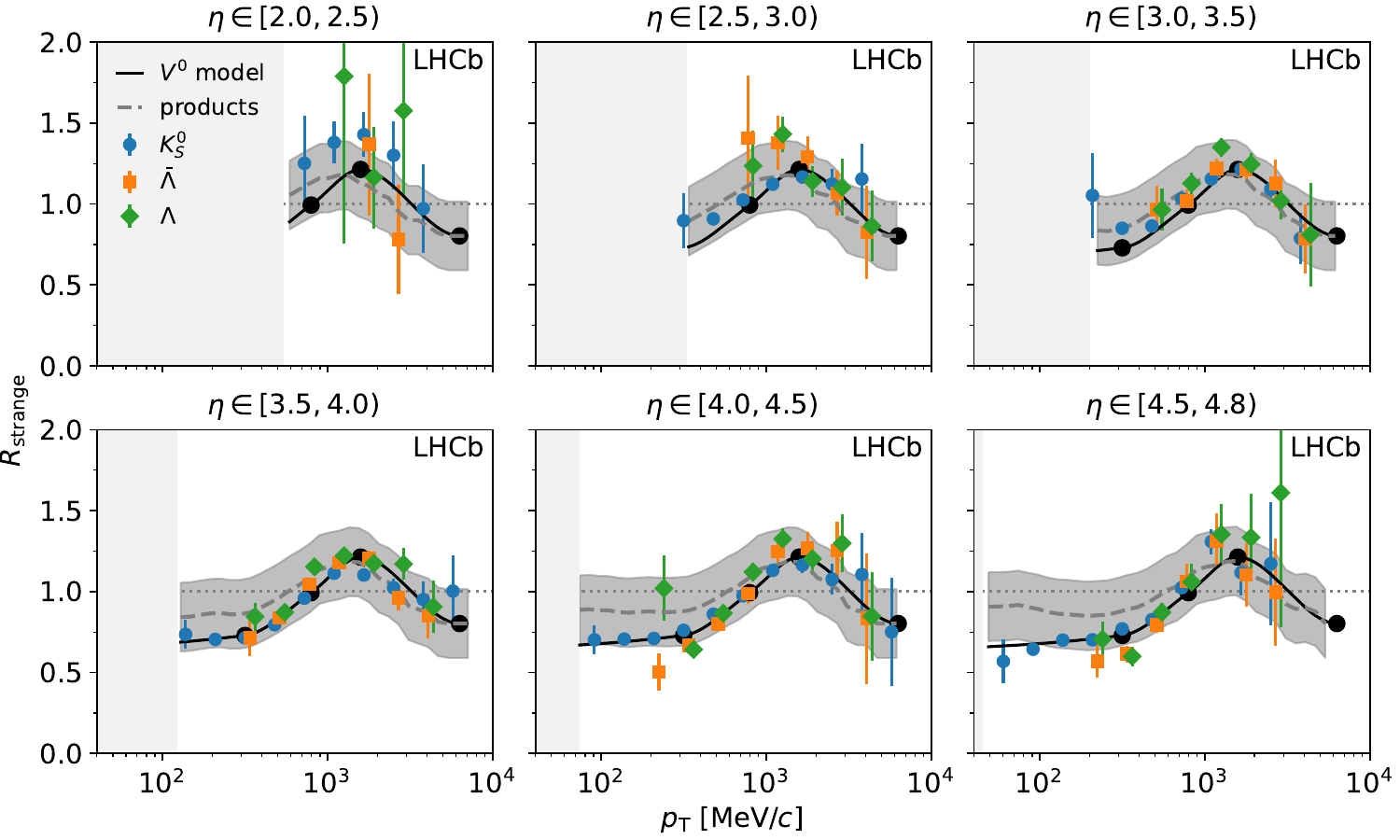}
  \caption{Ratio of the proxies for strange-hadron decays in data and simulation in intervals of $\eta$ and as a function of \pt for the \lhc fill recorded with the magnetic field pointing upwards and the simulated sample with the same magnetic-field configuration. The $\mathit{\Vzero}$~\textit{model} lines indicate the interpolated \KS, \Lz and \Lbar yield ratios, with support points indicated by black dots. The lines labelled as \textit{products} represent the proxy ratio of the decay products, with the bands representing the propagated systematic uncertainty. The light-grey areas indicate the limit of the kinematic acceptance.}
  \label{fig:v0_ratio_4201}
\end{figure}

The ratios of the signal yields in data and simulation are only computed in those kinematic intervals where a signal peak is present with a significance of at least three standard deviations. The ratios are shown in Fig.~\ref{fig:v0_ratio_4201}. In all $\eta$ intervals and for all three strange-hadron species, a pattern is observed in the yield ratios as a function of \pt. This is interpreted as a general difference in the amount of produced strange quarks between data and simulation, since it is the same for a meson and a baryon. The pattern is modelled with a monotone cubic spline~\cite{Fritsch:1980a}. A least-squares fit of the model to the yield ratios is performed, in which a Gaussian penalty term is added in order to restrict the value of the spline at the upper limit of the \pt range, where no data points are present.

The fitted model is then used to determine the proxy ratio of the strange hadrons, $R_\mathrm{parent}$, over the full kinematic range. However, the background for prompt long-lived charged particles is caused by the decay products of these hadrons. The proxy ratio of the products, $R_\mathrm{strange}$, in the $(\eta, \pt)$ intervals of their kinematic distributions is computed as
\begin{equation}
  R_{\mathrm{strange},\,k\ell} = \frac{\sum_{\hadron,\,ij} n_{\hadron,\,ijk\ell} \, R_{\mathrm{parent},\,j}}{\sum_{\hadron,\,ij} n_{\hadron,\,ijk\ell}} \, ,
\end{equation}
where: $\hadron$ is $\KS, \Lz, \Lbar$; $i$ and $j$ iterate through the $\eta$ and \pt intervals of the kinematic distributions of the parent particles, respectively; $k$ and $\ell$ iterate through the $\eta$ and \pt intervals of the kinematic distributions of the decay products, respectively; and $n$ is the simulated yield of the decay products in the corresponding intervals. The ratio $R_\mathrm{strange}$ is closer to unity than the ratio $R_\mathrm{parent}$, as the broad kinematic distributions of the decay products dilute deviations. The statistical uncertainty of the fitted model is negligible compared to the systematic uncertainty suggested by the deviations of the yield ratios from the model. A systematic uncertainty of $\pm 15\,\%$ is assigned to the proxy ratio to cover these deviations.

\section{Results}
\label{sec:results}

The differential cross-section of prompt inclusive production of long-lived charged particles is determined with Eqs.~\eqref{eqn:differential_cross_section} and~\eqref{eqn:particle_count}.

The statistical uncertainty of the data sample is subdominant in the full kinematic range, reaching $1.5\,\%$ at high \pt. Non-Poissonian statistical fluctuations of the candidate counts in kinematic intervals are taken into account. Each event contributes entries to a number of kinematic intervals, introducing statistical correlations of up to 0.4 between intervals, and variances that are up to $50\,\%$ larger than the Poisson expectation. The covariance matrix of these statistical fluctuations is estimated by dividing the full sample into 100 equivalent subsets, computing the covariance of the subsets and extrapolating the result back to the full data set.

\begin{figure}
  \centering
  \includegraphics[width=\textwidth]{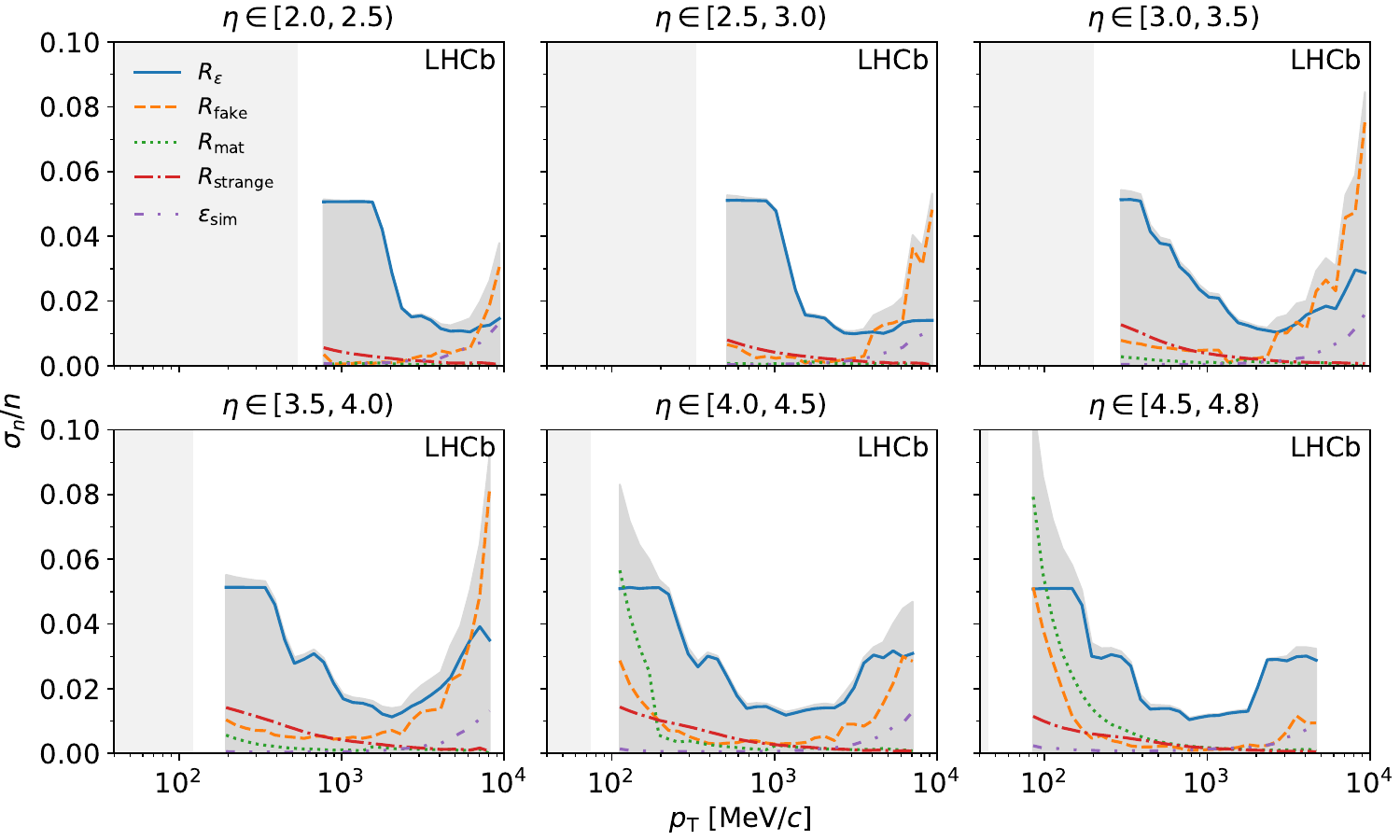}
  \caption{Leading sources of relative uncertainties of the number of prompt long-lived charged particles in intervals of $\eta$ and as a function of \pt for the \lhc fill recorded with the magnetic field pointing upwards. The total uncertainty, represented by the grey envelopes, excludes the uncertainty of the integrated luminosity of $2.0\,\%$. The light-grey areas indicate the limit of the kinematic acceptance.}
  \label{fig:final_error_simple_4201}
\end{figure}

The uncertainty on the differential cross-section is computed from the uncertainties of the individual terms in Eqs.~\eqref{eqn:differential_cross_section} and~\eqref{eqn:particle_count} using full error propagation of the respective covariance matrices. The contributions to the uncertainty of the number of prompt long-lived charged particles are shown in Fig.~\ref{fig:final_error_simple_4201}. The correction of the track-reconstruction efficiency is the largest contribution in most intervals. In the range ${\eta \in [2.0, 4.0)}$ and at high \pt, the uncertainty of the proxy for fake tracks dominates, while in the range ${\eta \in [4.0, 4.8)}$ and at low \pt, the uncertainty of the proxy for material interactions contributes most.

The final results are obtained by adding the fully corrected and background-subtracted numbers of prompt long-lived charged particles from both \lhc fills and dividing by the combined integrated luminosity, after confirming that the two separate results from each fill are consistent with each other. As a further cross-check, data from six other fills, recorded at the same centre-of-mass energy and with the same trigger as the default data sample, are compared and found to be consistent with the main result. For the calculation of the covariance matrix of the combined result, statistical uncertainties are treated as uncorrelated, while all systematic uncertainties, including the uncertainty of the luminosity, are taken to be fully correlated between the fills.

\begin{table}
  \centering
  \caption{Statistical and systematic uncertainties affecting the measured differential cross-section.}
  \label{tab:uncertainties}
  \begin{tabular}{l r@{--}l}
    \toprule
    Source & \multicolumn{2}{l}{Relative uncertainty in \%} \\
    \midrule
    Statistical uncertainty of the data sample & 0.0 & 1.5 \\
    Total efficiency & 0.9 & 5.1 \\
    Beam-gas interactions & 0.0 & 1.7 \\
    Fake tracks & 0.1 & 9.5 \\
    Material interactions & 0.0 & 12 \\
    Strange-hadron decays & 0.0 & 1.5 \\
    Other background contributions & 0.1 & 1.1 \\
    Integrated luminosity & \multicolumn{2}{l}{\hspace{4mm}2.0} \\
    \midrule
    Total uncertainty & 2.3 & 15 \\
    \bottomrule
  \end{tabular}
\end{table}

The minimum and maximum values over all kinematic intervals for the uncertainties from each source that contribute to the final result are listed in Table~\ref{tab:uncertainties}. The total uncertainty of the differential cross-section is between $2.3\,\%$ and $15\,\%$, which includes the fully correlated systematic uncertainty of $2.0\,\%$ from the integrated luminosity. The correlations are nonzero and positive between kinematic intervals and the two particle charges, since the systematic uncertainties dominate. The correlation matrix of the final differential cross-section is shown in Appendix~\ref{sec:correlation_matrix}.

\begin{figure}
  \centering
  \includegraphics[width=0.46\textwidth]{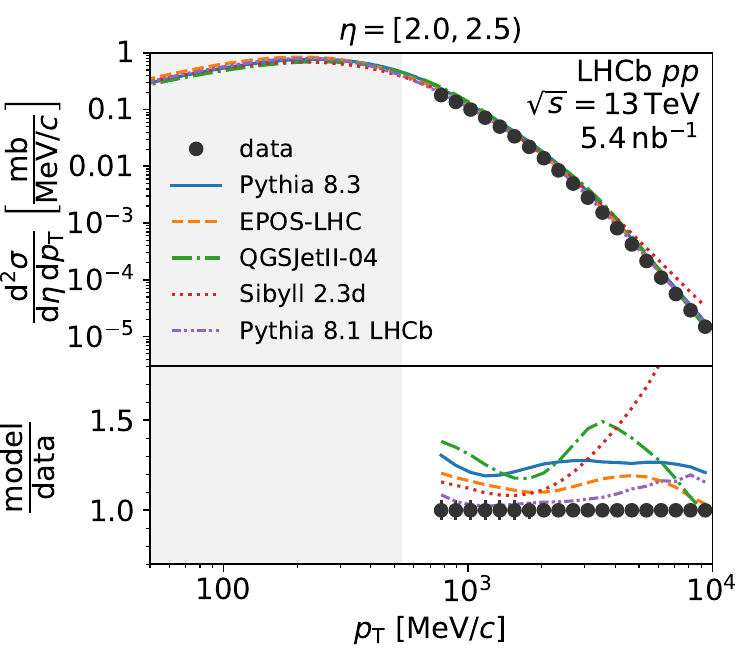}
  \includegraphics[width=0.46\textwidth]{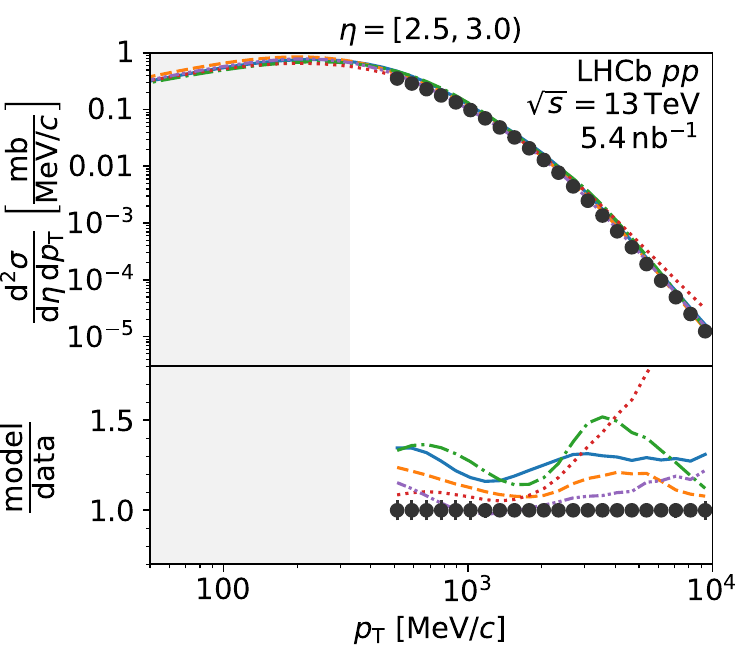}
  \includegraphics[width=0.46\textwidth]{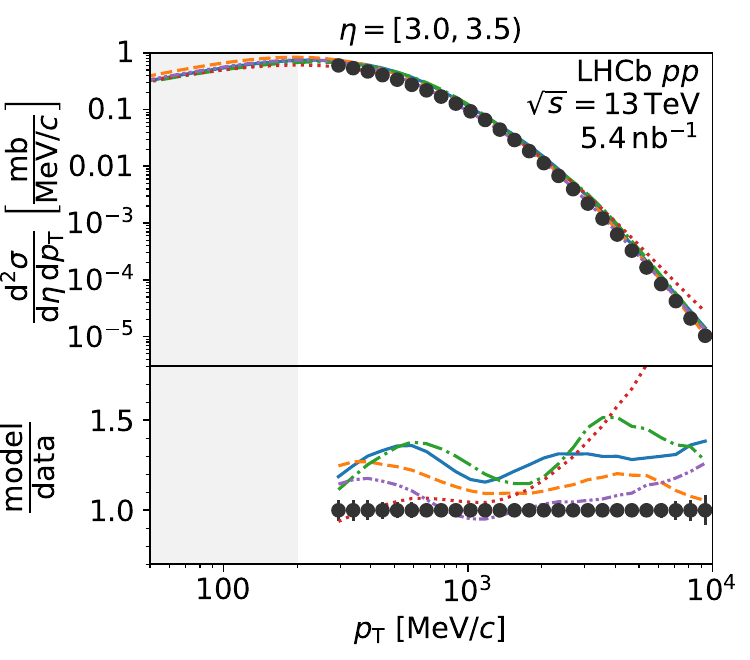}
  \includegraphics[width=0.46\textwidth]{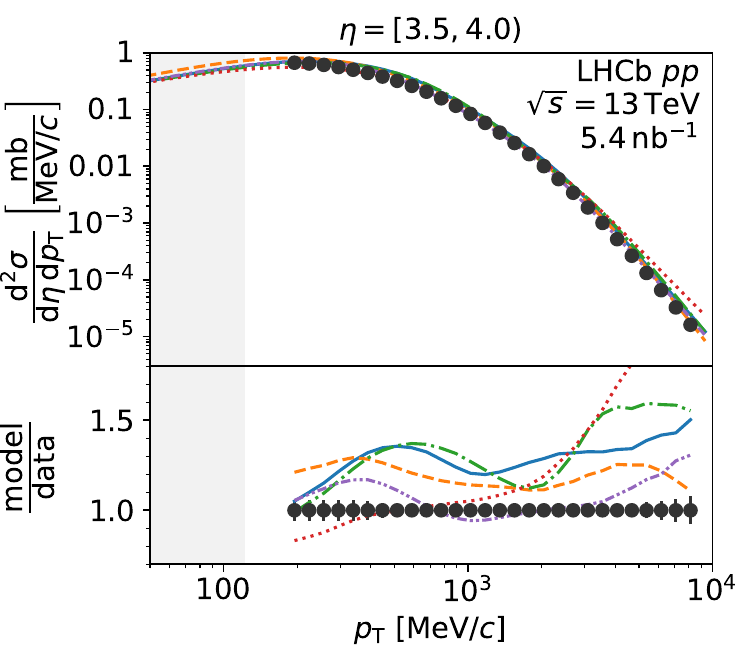}
  \includegraphics[width=0.46\textwidth]{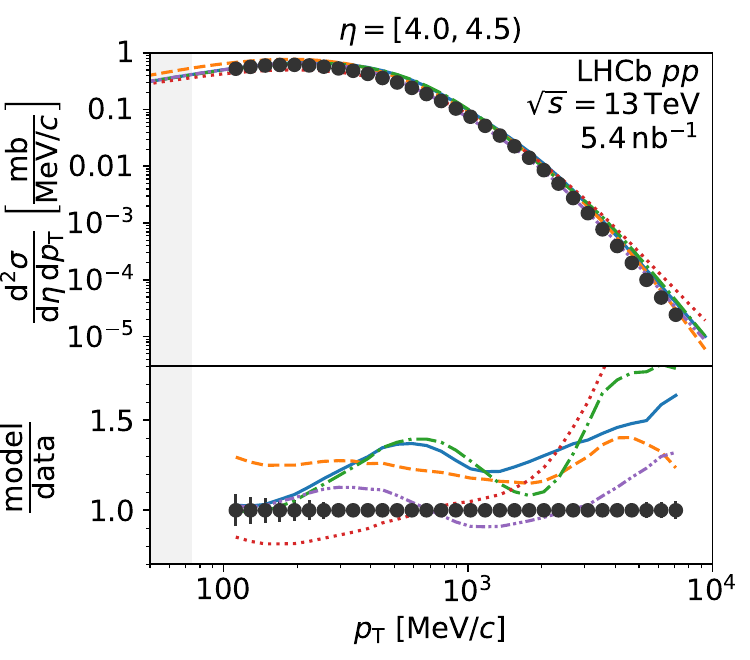}
  \includegraphics[width=0.46\textwidth]{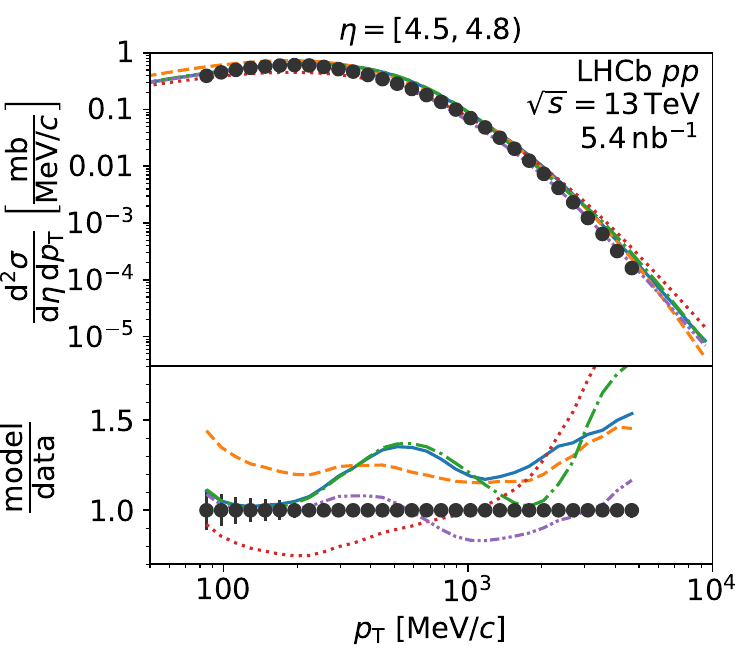}
  \caption{Differential cross-section of inclusive production of prompt long-lived charged particles in intervals of pseudorapidity, $\eta$, and as a function of transverse momentum, \pt. The error bars indicate the total uncertainty. The ratios of the model predictions and this measurement are shown in the lower panels. The lines labelled as \textit{Pythia~8.1~\lhcb} correspond to the occupancy-weighted simulated samples of this analysis.}
  \label{fig:final_0}
\end{figure}

\begin{figure}
  \centering
  \includegraphics[width=\textwidth]{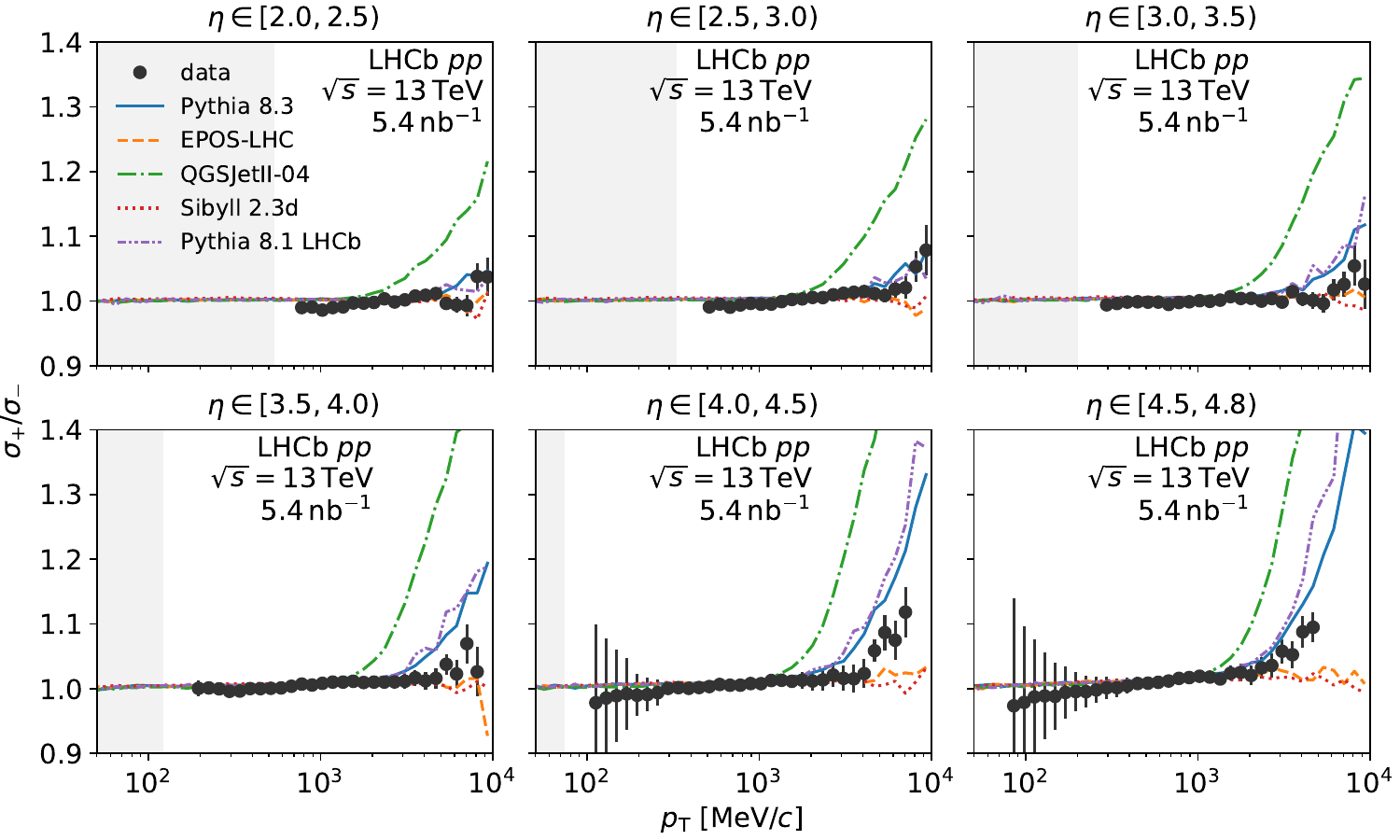}
  \caption{Ratios of the differential cross-sections of inclusive production of prompt long-lived positively and negatively charged particles as a function of transverse momentum and pseudorapidity for the data and the models shown in Fig.~\ref{fig:final_0}.}
  \label{fig:final_charge_ratio}
\end{figure}

\begin{table}
  \centering
  \caption{Inelastic cross-sections used in the models of which the predictions are compared with this measurement and values from recent measurements at the \lhc.}
  \label{tab:cross_sections}
  \begin{tabular}{l c}
    \toprule
    \multicolumn{1}{c}{Model or experiment} & {Inelastic cross-section in ${\!\mbarn}$} \\
    \midrule
    \pythia~8.3~\cite{Sjostrand:2014zea,*Sjostrand:2006za} & $78.05$ \\
    \textsc{EPOS-LHC}~\cite{Pierog:2013ria} & $78.98$ \\
    \textsc{QGSJet}~II-04~\cite{Ostapchenko:2010vb} & $80.17$ \\
    \textsc{Sibyll}~2.3d~\cite{Engel:2019dsg} & $79.86$ \\
    \midrule
    \atlas~\cite{Aaboud:2016mmw} & $78.1 \pm 2.9$ \\
    \lhcb~\cite{LHCb-PAPER-2018-003} & $75.4 \pm 5.4$ \\
    TOTEM~\cite{Antchev:2017dia} & $79.5 \pm 1.8$ \\
    \bottomrule
  \end{tabular}
\end{table}

The measured differential cross-section of inclusive production of prompt long-lived charged particles is shown in Fig.~\ref{fig:final_0}. The values for positively and negatively charged particles are listed in Appendix~\ref{sec:differential_cross_sections}. In the figure, the measurement is compared with the predictions from the hadronic-interaction models \pythia~8.3~\cite{Sjostrand:2014zea,*Sjostrand:2006za} with the default Monash tune~\cite{Skands:2014pea}, \textsc{EPOS-LHC}~\cite{Pierog:2013ria}, \textsc{QGSJet}~II-04~\cite{Ostapchenko:2010vb} and \textsc{Sibyll}~2.3d~\cite{Engel:2019dsg}. The latter three models are accessed through the \textsc{CRMC} package~\cite{Ulrich:2021a}. Also shown, but not comparable to the other models, is the occupancy-weighted \lhcb tune of \pythia~8.1 that was used as the simulated sample in this analysis. The calculation of the differential cross-section,
\begin{equation}
  \frac{\deriv^2 \sigma}{\deriv \eta \, \deriv \pt} \equiv \frac{\sigma_\mathrm{inel}}{N_\mathrm{inel}} \, \frac{n}{\Delta \eta \, \Delta \pt} \, ,
\end{equation}
is based on: the inelastic cross-section, $\sigma_\mathrm{inel}$, that is implemented in each model; the number of generated inelastic events, $N_\mathrm{inel}$; and the number of prompt long-lived charged particles, $n$, in each kinematic interval. The inelastic cross-sections of the models and recent measurements are listed in Table~\ref{tab:cross_sections}.

The deviations of the predictions from the measured values are between $-26\,\%$ and $+170\,\%$. The models mostly overestimate the differential cross-section. These deviations are caused by differences in the kinematic distributions compared to the data and a potential difference in the overall multiplicity of produced particles. The latter would cause an overall change in the magnitude of the differential cross-section. Differences in the inelastic cross-section have a similar effect, but can be excluded as the driving factor since the implemented values of the inelastic cross-section in the models are compatible with the measured ones. The occupancy-weighted \lhcb tune of \pythia~8.1 shows deviations between $-17\,\%$ and $+32\,\%$. The comparably good agreement of this simulation is based on the occupancy weighting, which allows only deviations in the shape, while for the remaining models, deviations in the magnitude can occur. Out of the other models, \textsc{EPOS-LHC} shows the smallest differences overall. The shape of the \pt distribution is well described; data and model differ mainly by an overall offset that weakly depends on $\eta$.

To address the question from the introduction, whether the pseudorapidity distribution is wide or narrow, the differential cross-section from \lhcb needs to be integrated over the \pt range and combined with the other measurements at lower and higher $\eta$ values. This combination is complicated by the fact that other experiments use (minimum-)biased triggers, while this analysis is unbiased. Therefore, this combination is not attempted in this paper. The \lhcb data do not cover the peak of the \pt distribution in all $\eta$ intervals. Since the peak contributes most to the integral, the integration of the \lhcb data is model dependent for ${\eta < 3.5}$.

The ratio of the differential cross-sections for positively and negatively charged particles is shown in Fig.~\ref{fig:final_charge_ratio}. The positively correlated components of the uncertainties cancel in the computation of the ratio. At ${\eta > 4.0}$ and low \pt, the uncertainty of the ratio is large due to the subtraction of background from material interactions, which cannot be assumed to be charge symmetric and which is not well constrained by control measurements in that region. At high $\eta$ and high \pt, the production of positively charged particles increases as the initial state has a charge of $+2$, which transfers to the final state in the forward region. This effect is predicted by the models to a varying extent. The best description of the ratio is provided by \pythia~8.3, although the onset of the increase is shifted towards lower \pt values. \textsc{QGSJet}~II-04 predicts an onset at even lower \pt values. \textsc{EPOS-LHC} and \textsc{Sibyll}~2.3d do not show a significant increase up to ${\pt = 10\,000\mevc}$, which is at tension with the measurement.

\section{Summary}
\label{sec:summary}

An unbiased measurement of the differential cross-section of prompt inclusive production of long-lived charged particles in \proton\proton collisions at ${\sqs = 13\tev}$ is presented. The data sample was recorded by the \lhcb experiment and corresponds to an integrated luminosity of $5.4\invnb$. The differential cross-section is measured as a function of transverse momentum and pseudorapidity in the ranges ${\pt \in [80, 10\,000)\mevc}$ and ${\eta \in [2.0, 4.8)}$ and is determined separately for positively and negatively charged particles. An uncertainty between $2.3\,\%$ and $15\,\%$, depending on \pt and $\eta$, is achieved. A comparison of the measured charge-combined differential cross-section with predictions from recent hadronic-interaction models shows that these models mostly overestimate the differential cross-section. The overall smallest deviations are observed for \textsc{EPOS-LHC}, while the ratio of the differential cross-sections for positively and negatively charged particles is best predicted by \pythia~8.3. The precision achieved in this measurement is essential for an improved simulation of the underlying event in collisions at the \lhc and of interactions in the atmosphere of the Earth that cause air showers.

\section*{Acknowledgements}
%
%
\noindent We express our gratitude to our colleagues in the CERN
accelerator departments for the excellent performance of the LHC. We
thank the technical and administrative staff at the LHCb
institutes.
We acknowledge support from CERN and from the national agencies:
CAPES, CNPq, FAPERJ and FINEP (Brazil);
MOST and NSFC (China);
CNRS/IN2P3 (France);
BMBF, DFG and MPG (Germany);
INFN (Italy);
NWO (Netherlands);
MNiSW and NCN (Poland);
MEN/IFA (Romania);
MSHE (Russia);
MICINN (Spain);
SNSF and SER (Switzerland);
NASU (Ukraine);
STFC (United Kingdom);
DOE NP and NSF (USA).
We acknowledge the computing resources that are provided by CERN, IN2P3
(France), KIT and DESY (Germany), INFN (Italy), SURF (Netherlands),
PIC (Spain), GridPP (United Kingdom), RRCKI and Yandex
LLC (Russia), CSCS (Switzerland), IFIN-HH (Romania), CBPF (Brazil),
PL-GRID (Poland) and NERSC (USA).
We are indebted to the communities behind the multiple open-source
software packages on which we depend.
Individual groups or members have received support from
ARC and ARDC (Australia);
AvH Foundation (Germany);
EPLANET, Marie Sk\l{}odowska-Curie Actions and ERC (European Union);
A*MIDEX, ANR, IPhU and Labex P2IO, and R\'{e}gion Auvergne-Rh\^{o}ne-Alpes (France);
Key Research Program of Frontier Sciences of CAS, CAS PIFI, CAS CCEPP,
Fundamental Research Funds for the Central Universities,
and Sci. \& Tech. Program of Guangzhou (China);
RFBR, RSF and Yandex LLC (Russia);
GVA, XuntaGal and GENCAT (Spain);
the Leverhulme Trust, the Royal Society
 and UKRI (United Kingdom).


\section*{Appendices}

\appendix

\section{Differential cross-sections}
\label{sec:differential_cross_sections}

The measured values of the differential cross-sections of prompt inclusive production of long-lived positively and negatively charged particles are listed in Table~\ref{tab:differential_cross_sections}. The numerical values and the correlation matrix are provided in machine-readable form at [link].

\begin{center}
\begin{longtable}{l l r r}
  \caption{Differential cross-sections of prompt inclusive production of long-lived positively and negatively charged particles as a function of transverse momentum and pseudorapidity.}
  \label{tab:differential_cross_sections} \\
  \toprule
  \multicolumn{1}{c}{$\eta$} & \multicolumn{1}{c}{$\log_{10}(\pt / (\!\gevc))$} & \multicolumn{2}{c}{$(\deriv^2 \sigma / (\deriv \eta \, \deriv \pt)) / (\!\mbarn / (\!\gevc))$} \\
  & & \multicolumn{1}{c}{$q = -1$} & \multicolumn{1}{c}{$q = +1$} \\
  \midrule
  \endfirsthead
  \midrule
  \multicolumn{1}{c}{$\eta$} & \multicolumn{1}{c}{$\log_{10}(\pt / (\!\gevc))$} & \multicolumn{2}{c}{$(\deriv^2 \sigma / (\deriv \eta \, \deriv \pt)) / (\!\mbarn / (\!\gevc))$} \\
  & & \multicolumn{1}{c}{$q = -1$} & \multicolumn{1}{c}{$q = +1$} \\
  \midrule
  \endhead
  $[2.0, 2.5)$ & $[-0.14, -0.08)$ & $90 \pm 5$ & $89 \pm 5$ \\
  $[2.0, 2.5)$ & $[-0.08, -0.02)$ & $68 \pm 4$ & $68 \pm 4$ \\
  $[2.0, 2.5)$ & $[-0.02, 0.04)$ & $50.6 \pm 2.8$ & $49.7 \pm 2.7$ \\
  $[2.0, 2.5)$ & $[0.04, 0.10)$ & $36.2 \pm 2.0$ & $35.7 \pm 2.0$ \\
  $[2.0, 2.5)$ & $[0.10, 0.16)$ & $25.2 \pm 1.4$ & $24.9 \pm 1.4$ \\
  $[2.0, 2.5)$ & $[0.16, 0.22)$ & $16.9 \pm 0.9$ & $16.8 \pm 0.9$ \\
  $[2.0, 2.5)$ & $[0.22, 0.28)$ & $11.0 \pm 0.5$ & $11.0 \pm 0.5$ \\
  $[2.0, 2.5)$ & $[0.28, 0.34)$ & $6.96 \pm 0.24$ & $6.94 \pm 0.24$ \\
  $[2.0, 2.5)$ & $[0.34, 0.40)$ & $4.23 \pm 0.12$ & $4.23 \pm 0.11$ \\
  $[2.0, 2.5)$ & $[0.40, 0.46)$ & $2.49 \pm 0.06$ & $2.48 \pm 0.06$ \\
  $[2.0, 2.5)$ & $[0.46, 0.52)$ & $1.40 \pm 0.04$ & $1.40 \pm 0.04$ \\
  $[2.0, 2.5)$ & $[0.52, 0.58)$ & $0.763 \pm 0.019$ & $0.766 \pm 0.019$ \\
  $[2.0, 2.5)$ & $[0.58, 0.64)$ & $0.405 \pm 0.010$ & $0.406 \pm 0.010$ \\
  $[2.0, 2.5)$ & $[0.64, 0.70)$ & $0.209 \pm 0.005$ & $0.212 \pm 0.005$ \\
  $[2.0, 2.5)$ & $[0.70, 0.76)$ & $0.1079 \pm 0.0026$ & $0.1069 \pm 0.0026$ \\
  $[2.0, 2.5)$ & $[0.76, 0.82)$ & $0.0550 \pm 0.0014$ & $0.0546 \pm 0.0014$ \\
  $[2.0, 2.5)$ & $[0.82, 0.88)$ & $(28.2 \pm 0.8) \cdot 10^{-3}$ & $(28.0 \pm 0.8) \cdot 10^{-3}$ \\
  $[2.0, 2.5)$ & $[0.88, 0.94)$ & $(14.3 \pm 0.5) \cdot 10^{-3}$ & $(14.7 \pm 0.5) \cdot 10^{-3}$ \\
  $[2.0, 2.5)$ & $[0.94, 1.00)$ & $(7.37 \pm 0.28) \cdot 10^{-3}$ & $(7.39 \pm 0.28) \cdot 10^{-3}$ \\
  $[2.5, 3.0)$ & $[-0.32, -0.26)$ & $177 \pm 10$ & $175 \pm 10$ \\
  $[2.5, 3.0)$ & $[-0.26, -0.20)$ & $144 \pm 8$ & $144 \pm 8$ \\
  $[2.5, 3.0)$ & $[-0.20, -0.14)$ & $115 \pm 6$ & $114 \pm 6$ \\
  $[2.5, 3.0)$ & $[-0.14, -0.08)$ & $89 \pm 5$ & $88 \pm 5$ \\
  $[2.5, 3.0)$ & $[-0.08, -0.02)$ & $67 \pm 4$ & $67 \pm 4$ \\
  $[2.5, 3.0)$ & $[-0.02, 0.04)$ & $49.1 \pm 2.6$ & $48.8 \pm 2.6$ \\
  $[2.5, 3.0)$ & $[0.04, 0.10)$ & $35.1 \pm 1.4$ & $34.9 \pm 1.4$ \\
  $[2.5, 3.0)$ & $[0.10, 0.16)$ & $24.3 \pm 0.8$ & $24.2 \pm 0.8$ \\
  $[2.5, 3.0)$ & $[0.16, 0.22)$ & $16.2 \pm 0.4$ & $16.2 \pm 0.4$ \\
  $[2.5, 3.0)$ & $[0.22, 0.28)$ & $10.40 \pm 0.27$ & $10.39 \pm 0.27$ \\
  $[2.5, 3.0)$ & $[0.28, 0.34)$ & $6.47 \pm 0.16$ & $6.47 \pm 0.16$ \\
  $[2.5, 3.0)$ & $[0.34, 0.40)$ & $3.87 \pm 0.09$ & $3.86 \pm 0.09$ \\
  $[2.5, 3.0)$ & $[0.40, 0.46)$ & $2.21 \pm 0.05$ & $2.22 \pm 0.05$ \\
  $[2.5, 3.0)$ & $[0.46, 0.52)$ & $1.236 \pm 0.028$ & $1.242 \pm 0.028$ \\
  $[2.5, 3.0)$ & $[0.52, 0.58)$ & $0.680 \pm 0.016$ & $0.681 \pm 0.016$ \\
  $[2.5, 3.0)$ & $[0.58, 0.64)$ & $0.358 \pm 0.009$ & $0.360 \pm 0.009$ \\
  $[2.5, 3.0)$ & $[0.64, 0.70)$ & $0.186 \pm 0.004$ & $0.189 \pm 0.005$ \\
  $[2.5, 3.0)$ & $[0.70, 0.76)$ & $0.0943 \pm 0.0026$ & $0.0967 \pm 0.0033$ \\
  $[2.5, 3.0)$ & $[0.76, 0.82)$ & $0.0481 \pm 0.0013$ & $0.0487 \pm 0.0015$ \\
  $[2.5, 3.0)$ & $[0.82, 0.88)$ & $(24.3 \pm 1.0) \cdot 10^{-3}$ & $0.0251 \pm 0.0012$ \\
  $[2.5, 3.0)$ & $[0.88, 0.94)$ & $(12.3 \pm 0.5) \cdot 10^{-3}$ & $(12.6 \pm 0.5) \cdot 10^{-3}$ \\
  $[2.5, 3.0)$ & $[0.94, 1.00)$ & $(6.04 \pm 0.34) \cdot 10^{-3}$ & $(6.5 \pm 0.4) \cdot 10^{-3}$ \\
  $[3.0, 3.5)$ & $[-0.56, -0.50)$ & $300 \pm 17$ & $299 \pm 17$ \\
  $[3.0, 3.5)$ & $[-0.50, -0.44)$ & $268 \pm 15$ & $268 \pm 15$ \\
  $[3.0, 3.5)$ & $[-0.44, -0.38)$ & $235 \pm 13$ & $234 \pm 13$ \\
  $[3.0, 3.5)$ & $[-0.38, -0.32)$ & $202 \pm 10$ & $202 \pm 10$ \\
  $[3.0, 3.5)$ & $[-0.32, -0.26)$ & $168 \pm 7$ & $168 \pm 7$ \\
  $[3.0, 3.5)$ & $[-0.26, -0.20)$ & $137 \pm 6$ & $137 \pm 6$ \\
  $[3.0, 3.5)$ & $[-0.20, -0.14)$ & $109 \pm 4$ & $109 \pm 4$ \\
  $[3.0, 3.5)$ & $[-0.14, -0.08)$ & $84.6 \pm 3.0$ & $84.6 \pm 3.0$ \\
  $[3.0, 3.5)$ & $[-0.08, -0.02)$ & $63.7 \pm 2.0$ & $63.7 \pm 2.0$ \\
  $[3.0, 3.5)$ & $[-0.02, 0.04)$ & $46.5 \pm 1.4$ & $46.5 \pm 1.4$ \\
  $[3.0, 3.5)$ & $[0.04, 0.10)$ & $32.8 \pm 1.0$ & $32.8 \pm 1.0$ \\
  $[3.0, 3.5)$ & $[0.10, 0.16)$ & $22.2 \pm 0.6$ & $22.2 \pm 0.6$ \\
  $[3.0, 3.5)$ & $[0.16, 0.22)$ & $14.45 \pm 0.35$ & $14.6 \pm 0.4$ \\
  $[3.0, 3.5)$ & $[0.22, 0.28)$ & $9.30 \pm 0.23$ & $9.29 \pm 0.23$ \\
  $[3.0, 3.5)$ & $[0.28, 0.34)$ & $5.71 \pm 0.14$ & $5.71 \pm 0.14$ \\
  $[3.0, 3.5)$ & $[0.34, 0.40)$ & $3.39 \pm 0.08$ & $3.40 \pm 0.08$ \\
  $[3.0, 3.5)$ & $[0.40, 0.46)$ & $1.98 \pm 0.05$ & $1.98 \pm 0.05$ \\
  $[3.0, 3.5)$ & $[0.46, 0.52)$ & $1.103 \pm 0.029$ & $1.102 \pm 0.027$ \\
  $[3.0, 3.5)$ & $[0.52, 0.58)$ & $0.597 \pm 0.018$ & $0.606 \pm 0.016$ \\
  $[3.0, 3.5)$ & $[0.58, 0.64)$ & $0.314 \pm 0.010$ & $0.314 \pm 0.008$ \\
  $[3.0, 3.5)$ & $[0.64, 0.70)$ & $0.163 \pm 0.006$ & $0.163 \pm 0.005$ \\
  $[3.0, 3.5)$ & $[0.70, 0.76)$ & $0.0830 \pm 0.0035$ & $0.0821 \pm 0.0029$ \\
  $[3.0, 3.5)$ & $[0.76, 0.82)$ & $0.0414 \pm 0.0017$ & $0.0422 \pm 0.0017$ \\
  $[3.0, 3.5)$ & $[0.82, 0.88)$ & $0.0207 \pm 0.0011$ & $0.0214 \pm 0.0011$ \\
  $[3.0, 3.5)$ & $[0.88, 0.94)$ & $(10.1 \pm 0.7) \cdot 10^{-3}$ & $(10.6 \pm 0.6) \cdot 10^{-3}$ \\
  $[3.0, 3.5)$ & $[0.94, 1.00)$ & $(5.2 \pm 0.5) \cdot 10^{-3}$ & $(5.1 \pm 0.4) \cdot 10^{-3}$ \\
  $[3.5, 4.0)$ & $[-0.74, -0.68)$ & $334 \pm 20$ & $335 \pm 20$ \\
  $[3.5, 4.0)$ & $[-0.68, -0.62)$ & $324 \pm 19$ & $324 \pm 19$ \\
  $[3.5, 4.0)$ & $[-0.62, -0.56)$ & $307 \pm 18$ & $306 \pm 18$ \\
  $[3.5, 4.0)$ & $[-0.56, -0.50)$ & $281 \pm 16$ & $280 \pm 16$ \\
  $[3.5, 4.0)$ & $[-0.50, -0.44)$ & $252 \pm 14$ & $251 \pm 14$ \\
  $[3.5, 4.0)$ & $[-0.44, -0.38)$ & $221 \pm 11$ & $221 \pm 11$ \\
  $[3.5, 4.0)$ & $[-0.38, -0.32)$ & $191 \pm 8$ & $191 \pm 8$ \\
  $[3.5, 4.0)$ & $[-0.32, -0.26)$ & $160 \pm 6$ & $161 \pm 6$ \\
  $[3.5, 4.0)$ & $[-0.26, -0.20)$ & $131 \pm 5$ & $131 \pm 5$ \\
  $[3.5, 4.0)$ & $[-0.20, -0.14)$ & $103 \pm 4$ & $103 \pm 4$ \\
  $[3.5, 4.0)$ & $[-0.14, -0.08)$ & $78.5 \pm 2.8$ & $79.3 \pm 2.8$ \\
  $[3.5, 4.0)$ & $[-0.08, -0.02)$ & $58.1 \pm 1.8$ & $58.6 \pm 1.8$ \\
  $[3.5, 4.0)$ & $[-0.02, 0.04)$ & $41.7 \pm 1.1$ & $42.1 \pm 1.2$ \\
  $[3.5, 4.0)$ & $[0.04, 0.10)$ & $29.0 \pm 0.8$ & $29.2 \pm 0.8$ \\
  $[3.5, 4.0)$ & $[0.10, 0.16)$ & $19.5 \pm 0.5$ & $19.8 \pm 0.5$ \\
  $[3.5, 4.0)$ & $[0.16, 0.22)$ & $12.80 \pm 0.33$ & $12.96 \pm 0.34$ \\
  $[3.5, 4.0)$ & $[0.22, 0.28)$ & $8.15 \pm 0.20$ & $8.23 \pm 0.21$ \\
  $[3.5, 4.0)$ & $[0.28, 0.34)$ & $5.02 \pm 0.13$ & $5.06 \pm 0.12$ \\
  $[3.5, 4.0)$ & $[0.34, 0.40)$ & $2.96 \pm 0.08$ & $2.99 \pm 0.07$ \\
  $[3.5, 4.0)$ & $[0.40, 0.46)$ & $1.70 \pm 0.05$ & $1.72 \pm 0.05$ \\
  $[3.5, 4.0)$ & $[0.46, 0.52)$ & $0.938 \pm 0.029$ & $0.945 \pm 0.026$ \\
  $[3.5, 4.0)$ & $[0.52, 0.58)$ & $0.498 \pm 0.016$ & $0.510 \pm 0.015$ \\
  $[3.5, 4.0)$ & $[0.58, 0.64)$ & $0.258 \pm 0.009$ & $0.262 \pm 0.008$ \\
  $[3.5, 4.0)$ & $[0.64, 0.70)$ & $0.132 \pm 0.005$ & $0.134 \pm 0.005$ \\
  $[3.5, 4.0)$ & $[0.70, 0.76)$ & $0.0647 \pm 0.0030$ & $0.0675 \pm 0.0030$ \\
  $[3.5, 4.0)$ & $[0.76, 0.82)$ & $0.0319 \pm 0.0017$ & $0.0337 \pm 0.0018$ \\
  $[3.5, 4.0)$ & $[0.82, 0.88)$ & $0.0159 \pm 0.0011$ & $0.0168 \pm 0.0010$ \\
  $[3.5, 4.0)$ & $[0.88, 0.94)$ & $(8.0 \pm 0.7) \cdot 10^{-3}$ & $(8.1 \pm 0.6) \cdot 10^{-3}$ \\
  $[4.0, 4.5)$ & $[-0.98, -0.92)$ & $264 \pm 28$ & $260 \pm 28$ \\
  $[4.0, 4.5)$ & $[-0.92, -0.86)$ & $285 \pm 26$ & $281 \pm 25$ \\
  $[4.0, 4.5)$ & $[-0.86, -0.80)$ & $300 \pm 23$ & $297 \pm 23$ \\
  $[4.0, 4.5)$ & $[-0.80, -0.74)$ & $306 \pm 22$ & $305 \pm 21$ \\
  $[4.0, 4.5)$ & $[-0.74, -0.68)$ & $308 \pm 19$ & $306 \pm 19$ \\
  $[4.0, 4.5)$ & $[-0.68, -0.62)$ & $301 \pm 17$ & $298 \pm 17$ \\
  $[4.0, 4.5)$ & $[-0.62, -0.56)$ & $286 \pm 14$ & $284 \pm 14$ \\
  $[4.0, 4.5)$ & $[-0.56, -0.50)$ & $265 \pm 10$ & $266 \pm 10$ \\
  $[4.0, 4.5)$ & $[-0.50, -0.44)$ & $241 \pm 8$ & $241 \pm 8$ \\
  $[4.0, 4.5)$ & $[-0.44, -0.38)$ & $213 \pm 8$ & $213 \pm 8$ \\
  $[4.0, 4.5)$ & $[-0.38, -0.32)$ & $180 \pm 6$ & $180 \pm 6$ \\
  $[4.0, 4.5)$ & $[-0.32, -0.26)$ & $150 \pm 5$ & $150 \pm 5$ \\
  $[4.0, 4.5)$ & $[-0.26, -0.20)$ & $120.2 \pm 3.3$ & $121.0 \pm 3.3$ \\
  $[4.0, 4.5)$ & $[-0.20, -0.14)$ & $93.5 \pm 2.4$ & $94.1 \pm 2.4$ \\
  $[4.0, 4.5)$ & $[-0.14, -0.08)$ & $70.5 \pm 1.8$ & $71.0 \pm 1.8$ \\
  $[4.0, 4.5)$ & $[-0.08, -0.02)$ & $51.9 \pm 1.3$ & $52.4 \pm 1.3$ \\
  $[4.0, 4.5)$ & $[-0.02, 0.04)$ & $37.2 \pm 0.9$ & $37.6 \pm 0.9$ \\
  $[4.0, 4.5)$ & $[0.04, 0.10)$ & $25.7 \pm 0.6$ & $26.0 \pm 0.6$ \\
  $[4.0, 4.5)$ & $[0.10, 0.16)$ & $17.3 \pm 0.4$ & $17.5 \pm 0.4$ \\
  $[4.0, 4.5)$ & $[0.16, 0.22)$ & $11.25 \pm 0.29$ & $11.37 \pm 0.28$ \\
  $[4.0, 4.5)$ & $[0.22, 0.28)$ & $7.03 \pm 0.18$ & $7.13 \pm 0.18$ \\
  $[4.0, 4.5)$ & $[0.28, 0.34)$ & $4.26 \pm 0.12$ & $4.30 \pm 0.11$ \\
  $[4.0, 4.5)$ & $[0.34, 0.40)$ & $2.47 \pm 0.07$ & $2.49 \pm 0.06$ \\
  $[4.0, 4.5)$ & $[0.40, 0.46)$ & $1.36 \pm 0.04$ & $1.40 \pm 0.04$ \\
  $[4.0, 4.5)$ & $[0.46, 0.52)$ & $0.741 \pm 0.025$ & $0.748 \pm 0.021$ \\
  $[4.0, 4.5)$ & $[0.52, 0.58)$ & $0.386 \pm 0.015$ & $0.390 \pm 0.014$ \\
  $[4.0, 4.5)$ & $[0.58, 0.64)$ & $0.195 \pm 0.008$ & $0.200 \pm 0.007$ \\
  $[4.0, 4.5)$ & $[0.64, 0.70)$ & $0.096 \pm 0.004$ & $0.103 \pm 0.004$ \\
  $[4.0, 4.5)$ & $[0.70, 0.76)$ & $0.0478 \pm 0.0022$ & $0.0523 \pm 0.0024$ \\
  $[4.0, 4.5)$ & $[0.76, 0.82)$ & $0.0235 \pm 0.0011$ & $0.0253 \pm 0.0013$ \\
  $[4.0, 4.5)$ & $[0.82, 0.88)$ & $(11.3 \pm 0.6) \cdot 10^{-3}$ & $(13.0 \pm 0.7) \cdot 10^{-3}$ \\
  $[4.5, 4.8)$ & $[-1.10, -1.04)$ & $199 \pm 28$ & $195 \pm 27$ \\
  $[4.5, 4.8)$ & $[-1.04, -0.98)$ & $228 \pm 25$ & $224 \pm 24$ \\
  $[4.5, 4.8)$ & $[-0.98, -0.92)$ & $251 \pm 23$ & $249 \pm 22$ \\
  $[4.5, 4.8)$ & $[-0.92, -0.86)$ & $272 \pm 21$ & $270 \pm 20$ \\
  $[4.5, 4.8)$ & $[-0.86, -0.80)$ & $289 \pm 20$ & $284 \pm 19$ \\
  $[4.5, 4.8)$ & $[-0.80, -0.74)$ & $299 \pm 18$ & $298 \pm 17$ \\
  $[4.5, 4.8)$ & $[-0.74, -0.68)$ & $304 \pm 13$ & $303 \pm 13$ \\
  $[4.5, 4.8)$ & $[-0.68, -0.62)$ & $300 \pm 12$ & $299 \pm 12$ \\
  $[4.5, 4.8)$ & $[-0.62, -0.56)$ & $282 \pm 11$ & $282 \pm 11$ \\
  $[4.5, 4.8)$ & $[-0.56, -0.50)$ & $258 \pm 10$ & $258 \pm 10$ \\
  $[4.5, 4.8)$ & $[-0.50, -0.44)$ & $233 \pm 8$ & $233 \pm 8$ \\
  $[4.5, 4.8)$ & $[-0.44, -0.38)$ & $203 \pm 6$ & $204 \pm 6$ \\
  $[4.5, 4.8)$ & $[-0.38, -0.32)$ & $171 \pm 4$ & $172 \pm 4$ \\
  $[4.5, 4.8)$ & $[-0.32, -0.26)$ & $142 \pm 4$ & $143 \pm 4$ \\
  $[4.5, 4.8)$ & $[-0.26, -0.20)$ & $114.8 \pm 2.9$ & $115.9 \pm 2.9$ \\
  $[4.5, 4.8)$ & $[-0.20, -0.14)$ & $89.3 \pm 2.2$ & $90.4 \pm 2.2$ \\
  $[4.5, 4.8)$ & $[-0.14, -0.08)$ & $67.7 \pm 1.6$ & $68.4 \pm 1.6$ \\
  $[4.5, 4.8)$ & $[-0.08, -0.02)$ & $49.6 \pm 1.2$ & $50.5 \pm 1.2$ \\
  $[4.5, 4.8)$ & $[-0.02, 0.04)$ & $35.0 \pm 0.8$ & $35.7 \pm 0.8$ \\
  $[4.5, 4.8)$ & $[0.04, 0.10)$ & $24.0 \pm 0.6$ & $24.4 \pm 0.6$ \\
  $[4.5, 4.8)$ & $[0.10, 0.16)$ & $15.8 \pm 0.4$ & $16.1 \pm 0.4$ \\
  $[4.5, 4.8)$ & $[0.16, 0.22)$ & $10.08 \pm 0.25$ & $10.34 \pm 0.25$ \\
  $[4.5, 4.8)$ & $[0.22, 0.28)$ & $6.22 \pm 0.16$ & $6.35 \pm 0.15$ \\
  $[4.5, 4.8)$ & $[0.28, 0.34)$ & $3.66 \pm 0.12$ & $3.72 \pm 0.11$ \\
  $[4.5, 4.8)$ & $[0.34, 0.40)$ & $2.05 \pm 0.08$ & $2.12 \pm 0.07$ \\
  $[4.5, 4.8)$ & $[0.40, 0.46)$ & $1.14 \pm 0.04$ & $1.18 \pm 0.04$ \\
  $[4.5, 4.8)$ & $[0.46, 0.52)$ & $0.596 \pm 0.023$ & $0.631 \pm 0.022$ \\
  $[4.5, 4.8)$ & $[0.52, 0.58)$ & $0.314 \pm 0.013$ & $0.328 \pm 0.012$ \\
  $[4.5, 4.8)$ & $[0.58, 0.64)$ & $0.152 \pm 0.006$ & $0.169 \pm 0.006$ \\
  $[4.5, 4.8)$ & $[0.64, 0.70)$ & $0.075 \pm 0.004$ & $0.0854 \pm 0.0033$\\
  \bottomrule
\end{longtable}
\end{center}

\section{Correlation matrix of differential cross-section}
\label{sec:correlation_matrix}

\begin{figure}[b]
  \centering
  \includegraphics[width=\textwidth]{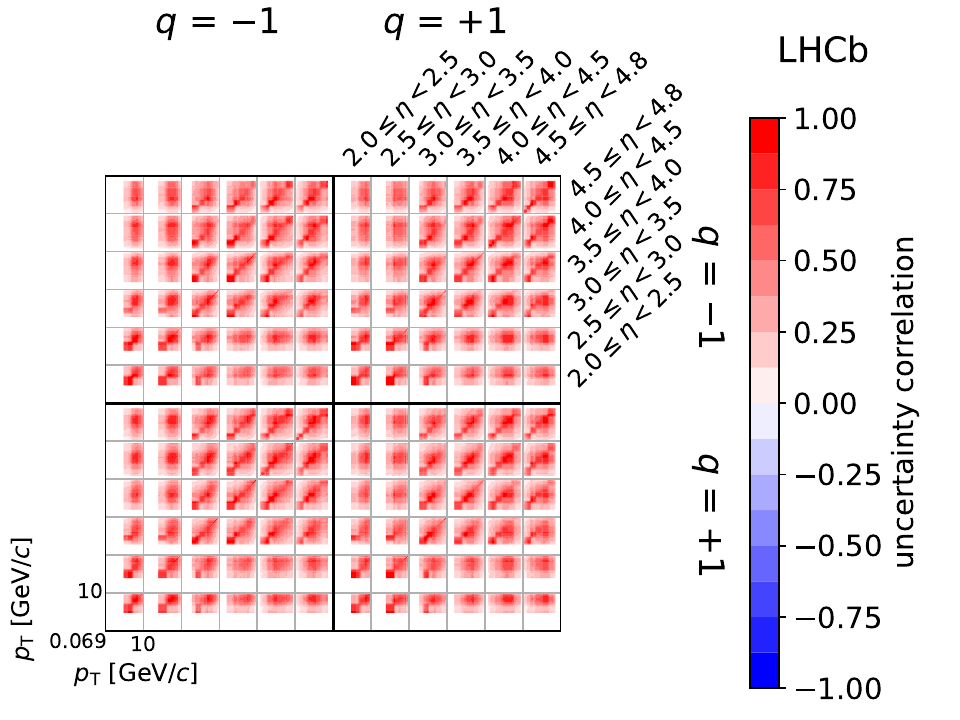}
  \caption{Correlation matrix for the uncertainties of the differential cross-section of prompt inclusive production of long-lived charged particles. The four large quadrants correspond to the correlations between negatively and positively charged particles. The 36 cells within each quadrant correspond to the $\eta$ intervals. In each cell, the correlations of the \pt intervals are shown from low to high \pt in logarithmic intervals.}
  \label{fig:final_correlation}
\end{figure}

The correlation matrix of the measured differential cross-section is shown in Fig.~\ref{fig:final_correlation}. The correlations are positive since the systematic uncertainties dominate and most of the systematic uncertainties are positively correlated between kinematic intervals and the two particle charges. The positive correlations between neighbouring \pt intervals are particularly large due to the correction of the tracking efficiency, which affects neighbouring intervals in the same way. The numerical values of the correlation matrix are provided in machine-readable form at [link].


\addcontentsline{toc}{section}{References}
\bibliographystyle{LHCb}
\bibliography{main,standard,LHCb-PAPER,LHCb-CONF,LHCb-DP,LHCb-TDR}

\ifx\mcitethebibliography\mciteundefinedmacro
\PackageError{LHCb.bst}{mciteplus.sty has not been loaded}
{This bibstyle requires the use of the mciteplus package.}\fi
\providecommand{\href}[2]{#2}
\begin{mcitethebibliography}{10}
\mciteSetBstSublistMode{n}
\mciteSetBstMaxWidthForm{subitem}{\alph{mcitesubitemcount})}
\mciteSetBstSublistLabelBeginEnd{\mcitemaxwidthsubitemform\space}
{\relax}{\relax}

\bibitem{Low:1975sv}
F.~E. Low, \ifthenelse{\boolean{articletitles}}{\emph{{Model of the bare
  pomeron}}, }{}\href{https://doi.org/10.1103/PhysRevD.12.163}{Phys.\ Rev.\
  \textbf{D12} (1975) 163}\relax
\mciteBstWouldAddEndPuncttrue
\mciteSetBstMidEndSepPunct{\mcitedefaultmidpunct}
{\mcitedefaultendpunct}{\mcitedefaultseppunct}\relax
\EndOfBibitem
\bibitem{Mueller:1986ey}
A.~H. Mueller and H.~Navelet, \ifthenelse{\boolean{articletitles}}{\emph{{An
  inclusive minijet cross section and the bare pomeron in QCD}},
  }{}\href{https://doi.org/10.1016/0550-3213(87)90705-X}{Nucl.\ Phys.\
  \textbf{B282} (1987) 727}\relax
\mciteBstWouldAddEndPuncttrue
\mciteSetBstMidEndSepPunct{\mcitedefaultmidpunct}
{\mcitedefaultendpunct}{\mcitedefaultseppunct}\relax
\EndOfBibitem
\bibitem{Nikolaev:1991et}
N.~N. Nikolaev and B.~G. Zakharov,
  \ifthenelse{\boolean{articletitles}}{\emph{{Pomeron structure function and
  diffraction dissociation of virtual photons in perturbative QCD}},
  }{}\href{https://doi.org/10.1007/BF01597573}{Z.\ Phys.\  \textbf{C53} (1992)
  331}\relax
\mciteBstWouldAddEndPuncttrue
\mciteSetBstMidEndSepPunct{\mcitedefaultmidpunct}
{\mcitedefaultendpunct}{\mcitedefaultseppunct}\relax
\EndOfBibitem
\bibitem{Goulianos:1994ph}
K.~Goulianos, \ifthenelse{\boolean{articletitles}}{\emph{{Universality of
  particle multiplicities}}, }{}
  {\href{https://www.osti.gov/biblio/10186266}{3rd Gleb Wataghin School on High
  Energy Phenomenology}}, 1994\relax
\mciteBstWouldAddEndPuncttrue
\mciteSetBstMidEndSepPunct{\mcitedefaultmidpunct}
{\mcitedefaultendpunct}{\mcitedefaultseppunct}\relax
\EndOfBibitem
\bibitem{Drescher:2000ha}
H.~J. Drescher {\em et~al.},
  \ifthenelse{\boolean{articletitles}}{\emph{{Parton-based Gribov--Regge
  theory}}, }{}\href{https://doi.org/10.1016/S0370-1573(00)00122-8}{Phys.\
  Rep.\  \textbf{350} (2001) 93},
  \href{http://arxiv.org/abs/hep-ph/0007198}{{\normalfont\ttfamily
  arXiv:hep-ph/0007198}}\relax
\mciteBstWouldAddEndPuncttrue
\mciteSetBstMidEndSepPunct{\mcitedefaultmidpunct}
{\mcitedefaultendpunct}{\mcitedefaultseppunct}\relax
\EndOfBibitem
\bibitem{Dembinski:2019uta}
EAS-MSU, IceCube, KASCADE-Grande, NEVOD-DECOR, Pierre Auger, SUGAR, Telescope
  Array and Yakutsk EAS Array collaborations, H.~P. Dembinski {\em et~al.},
  \ifthenelse{\boolean{articletitles}}{\emph{{Report on tests and measurements
  of hadronic interaction properties with air showers}},
  }{}\href{https://doi.org/10.1051/epjconf/201921002004}{EPJ Web Conf.\
  \textbf{210} (2019) 02004},
  \href{http://arxiv.org/abs/1902.08124}{{\normalfont\ttfamily
  arXiv:1902.08124}}\relax
\mciteBstWouldAddEndPuncttrue
\mciteSetBstMidEndSepPunct{\mcitedefaultmidpunct}
{\mcitedefaultendpunct}{\mcitedefaultseppunct}\relax
\EndOfBibitem
\bibitem{Dembinski:2020flw}
H.~P. Dembinski, \ifthenelse{\boolean{articletitles}}{\emph{{The Muon Puzzle in
  high-energy air showers}},
  }{}\href{https://doi.org/10.1134/S1063778819660165}{Phys.\ Atom.\ Nuclei
  \textbf{82} (2019) 644}\relax
\mciteBstWouldAddEndPuncttrue
\mciteSetBstMidEndSepPunct{\mcitedefaultmidpunct}
{\mcitedefaultendpunct}{\mcitedefaultseppunct}\relax
\EndOfBibitem
\bibitem{Citron:2018lsq}
Z.~Citron {\em et~al.}, \ifthenelse{\boolean{articletitles}}{\emph{{Future
  physics opportunities for high-density QCD at the \lhc with heavy-ion and
  proton beams}}, }{}\href{https://doi.org/10.23731/CYRM-2019-007.1159}{\cern
  Yellow Rep.\ Monogr.\  \textbf{7} (2019) 1159},
  \href{http://arxiv.org/abs/1812.06772}{{\normalfont\ttfamily
  arXiv:1812.06772}}\relax
\mciteBstWouldAddEndPuncttrue
\mciteSetBstMidEndSepPunct{\mcitedefaultmidpunct}
{\mcitedefaultendpunct}{\mcitedefaultseppunct}\relax
\EndOfBibitem
\bibitem{Baur:2019cpv}
S.~Baur {\em et~al.}, \ifthenelse{\boolean{articletitles}}{\emph{{Core-corona
  effect in hadron collisions and muon production in air showers}},
  }{}\href{http://arxiv.org/abs/1902.09265}{{\normalfont\ttfamily
  arXiv:1902.09265}}\relax
\mciteBstWouldAddEndPuncttrue
\mciteSetBstMidEndSepPunct{\mcitedefaultmidpunct}
{\mcitedefaultendpunct}{\mcitedefaultseppunct}\relax
\EndOfBibitem
\bibitem{Albrecht:2021yla}
J.~Albrecht {\em et~al.}, \ifthenelse{\boolean{articletitles}}{\emph{{The Muon
  Puzzle in cosmic-ray induced air showers and its connection to the Large
  Hadron Collider}},
  }{}\href{http://arxiv.org/abs/2105.06148}{{\normalfont\ttfamily
  arXiv:2105.06148}}\relax
\mciteBstWouldAddEndPuncttrue
\mciteSetBstMidEndSepPunct{\mcitedefaultmidpunct}
{\mcitedefaultendpunct}{\mcitedefaultseppunct}\relax
\EndOfBibitem
\bibitem{Acharya:2017a}
\alice collaboration, S.~Acharya {\em et~al.},
  \ifthenelse{\boolean{articletitles}}{\emph{{The \alice definition of primary
  particles}}, }{}
  {\href{https://cds.cern.ch/record/2270008}{ALICE-PUBLIC-2017-005}},
  2017\relax
\mciteBstWouldAddEndPuncttrue
\mciteSetBstMidEndSepPunct{\mcitedefaultmidpunct}
{\mcitedefaultendpunct}{\mcitedefaultseppunct}\relax
\EndOfBibitem
\bibitem{Engel:2019dsg}
F.~Riehn {\em et~al.}, \ifthenelse{\boolean{articletitles}}{\emph{{Hadronic
  interaction model Sibyll~2.3d and extensive air showers}},
  }{}\href{https://doi.org/10.1103/PhysRevD.102.063002}{Phys.\ Rev.\
  \textbf{D102} (2020) 063002},
  \href{http://arxiv.org/abs/1912.03300}{{\normalfont\ttfamily
  arXiv:1912.03300}}\relax
\mciteBstWouldAddEndPuncttrue
\mciteSetBstMidEndSepPunct{\mcitedefaultmidpunct}
{\mcitedefaultendpunct}{\mcitedefaultseppunct}\relax
\EndOfBibitem
\bibitem{Roesler:2000he}
S.~Roesler, R.~Engel, and J.~Ranft,
  \ifthenelse{\boolean{articletitles}}{\emph{{The Monte Carlo event generator
  DPMJET-III}},
  }{}\href{http://arxiv.org/abs/hep-ph/0012252}{{\normalfont\ttfamily
  arXiv:hep-ph/0012252}}\relax
\mciteBstWouldAddEndPuncttrue
\mciteSetBstMidEndSepPunct{\mcitedefaultmidpunct}
{\mcitedefaultendpunct}{\mcitedefaultseppunct}\relax
\EndOfBibitem
\bibitem{Fedynitch:2015kcn}
A.~Fedynitch, {\em {Cascade equations and hadronic interactions at very high
  energies}}, PhD thesis, Karlsruhe Institute of Technology, Karlsruhe, 2015,
  {\href{https://doi.org/10.5445/IR/1000055433}{CERN-THESIS-2015-371}}\relax
\mciteBstWouldAddEndPuncttrue
\mciteSetBstMidEndSepPunct{\mcitedefaultmidpunct}
{\mcitedefaultendpunct}{\mcitedefaultseppunct}\relax
\EndOfBibitem
\bibitem{Pierog:2013ria}
T.~Pierog {\em et~al.}, \ifthenelse{\boolean{articletitles}}{\emph{{EPOS LHC:
  Test of collective hadronization with data measured at the \cern Large Hadron
  Collider}}, }{}\href{https://doi.org/10.1103/PhysRevC.92.034906}{Phys.\ Rev.\
   \textbf{C92} (2015) 034906},
  \href{http://arxiv.org/abs/1306.0121}{{\normalfont\ttfamily
  arXiv:1306.0121}}\relax
\mciteBstWouldAddEndPuncttrue
\mciteSetBstMidEndSepPunct{\mcitedefaultmidpunct}
{\mcitedefaultendpunct}{\mcitedefaultseppunct}\relax
\EndOfBibitem
\bibitem{Ostapchenko:2010vb}
S.~Ostapchenko, \ifthenelse{\boolean{articletitles}}{\emph{{Monte Carlo
  treatment of hadronic interactions in enhanced pomeron scheme: QGSJET-II
  model}}, }{}\href{https://doi.org/10.1103/PhysRevD.83.014018}{Phys.\ Rev.\
  \textbf{D83} (2011) 014018},
  \href{http://arxiv.org/abs/1010.1869}{{\normalfont\ttfamily
  arXiv:1010.1869}}\relax
\mciteBstWouldAddEndPuncttrue
\mciteSetBstMidEndSepPunct{\mcitedefaultmidpunct}
{\mcitedefaultendpunct}{\mcitedefaultseppunct}\relax
\EndOfBibitem
\bibitem{Sjostrand:2014zea}
T.~Sj\"ostrand {\em et~al.}, \ifthenelse{\boolean{articletitles}}{\emph{{An
  introduction to PYTHIA 8.2}},
  }{}\href{https://doi.org/10.1016/j.cpc.2015.01.024}{Comput.\ Phys.\ Commun.\
  \textbf{191} (2015) 159},
  \href{http://arxiv.org/abs/1410.3012}{{\normalfont\ttfamily
  arXiv:1410.3012}}\relax
\mciteBstWouldAddEndPuncttrue
\mciteSetBstMidEndSepPunct{\mcitedefaultmidpunct}
{\mcitedefaultendpunct}{\mcitedefaultseppunct}\relax
\EndOfBibitem
\bibitem{Sjostrand:2006za}
T.~Sj\"{o}strand, S.~Mrenna, and P.~Skands,
  \ifthenelse{\boolean{articletitles}}{\emph{{PYTHIA 6.4 physics and manual}},
  }{}\href{https://doi.org/10.1088/1126-6708/2006/05/026}{JHEP \textbf{05}
  (2006) 026}, \href{http://arxiv.org/abs/hep-ph/0603175}{{\normalfont\ttfamily
  arXiv:hep-ph/0603175}}\relax
\mciteBstWouldAddEndPuncttrue
\mciteSetBstMidEndSepPunct{\mcitedefaultmidpunct}
{\mcitedefaultendpunct}{\mcitedefaultseppunct}\relax
\EndOfBibitem
\bibitem{ALICE:2015qqj}
\alice collaboration, J.~Adam {\em et~al.},
  \ifthenelse{\boolean{articletitles}}{\emph{{Pseudorapidity and
  transverse-momentum distributions of charged particles in proton--proton
  collisions at ${\sqs = 13\tev}$}},
  }{}\href{https://doi.org/10.1016/j.physletb.2015.12.030}{Phys.\ Lett.\
  \textbf{B753} (2016) 319},
  \href{http://arxiv.org/abs/1509.08734}{{\normalfont\ttfamily
  arXiv:1509.08734}}\relax
\mciteBstWouldAddEndPuncttrue
\mciteSetBstMidEndSepPunct{\mcitedefaultmidpunct}
{\mcitedefaultendpunct}{\mcitedefaultseppunct}\relax
\EndOfBibitem
\bibitem{ALICE:2018vuu}
\alice collaboration, S.~Acharya {\em et~al.},
  \ifthenelse{\boolean{articletitles}}{\emph{{Transverse momentum spectra and
  nuclear modification factors of charged particles in pp, p-Pb and Pb-Pb
  collisions at the \lhc}},
  }{}\href{https://doi.org/10.1007/JHEP11(2018)013}{JHEP \textbf{11} (2018)
  013}, \href{http://arxiv.org/abs/1802.09145}{{\normalfont\ttfamily
  arXiv:1802.09145}}\relax
\mciteBstWouldAddEndPuncttrue
\mciteSetBstMidEndSepPunct{\mcitedefaultmidpunct}
{\mcitedefaultendpunct}{\mcitedefaultseppunct}\relax
\EndOfBibitem
\bibitem{ATLAS:2010jvh}
\atlas collaboration, G.~Aad {\em et~al.},
  \ifthenelse{\boolean{articletitles}}{\emph{{Charged-particle multiplicities
  in \proton\proton interactions measured with the \atlas detector at the
  \lhc}}, }{}\href{https://doi.org/10.1088/1367-2630/13/5/053033}{New J.\
  Phys.\  \textbf{13} (2011) 053033},
  \href{http://arxiv.org/abs/1012.5104}{{\normalfont\ttfamily
  arXiv:1012.5104}}\relax
\mciteBstWouldAddEndPuncttrue
\mciteSetBstMidEndSepPunct{\mcitedefaultmidpunct}
{\mcitedefaultendpunct}{\mcitedefaultseppunct}\relax
\EndOfBibitem
\bibitem{ATLAS:2016qux}
\atlas collaboration, G.~Aad {\em et~al.},
  \ifthenelse{\boolean{articletitles}}{\emph{{Charged-particle distributions in
  \proton\proton interactions at ${\sqs = 8\tev}$ measured with the \atlas
  detector}}, }{}\href{https://doi.org/10.1140/epjc/s10052-016-4203-9}{Eur.\
  Phys.\ J.\  \textbf{C76} (2016) 403},
  \href{http://arxiv.org/abs/1603.02439}{{\normalfont\ttfamily
  arXiv:1603.02439}}\relax
\mciteBstWouldAddEndPuncttrue
\mciteSetBstMidEndSepPunct{\mcitedefaultmidpunct}
{\mcitedefaultendpunct}{\mcitedefaultseppunct}\relax
\EndOfBibitem
\bibitem{ATLAS:2016zkp}
\atlas collaboration, G.~Aad {\em et~al.},
  \ifthenelse{\boolean{articletitles}}{\emph{{Charged-particle distributions in
  ${\sqs = 13\tev}$ \proton\proton interactions measured with the \atlas
  detector at the \lhc}},
  }{}\href{https://doi.org/10.1016/j.physletb.2016.04.050}{Phys.\ Lett.\
  \textbf{B758} (2016) 67},
  \href{http://arxiv.org/abs/1602.01633}{{\normalfont\ttfamily
  arXiv:1602.01633}}\relax
\mciteBstWouldAddEndPuncttrue
\mciteSetBstMidEndSepPunct{\mcitedefaultmidpunct}
{\mcitedefaultendpunct}{\mcitedefaultseppunct}\relax
\EndOfBibitem
\bibitem{ATLAS:2016zba}
\atlas collaboration, M.~Aaboud {\em et~al.},
  \ifthenelse{\boolean{articletitles}}{\emph{{Charged-particle distributions at
  low transverse momentum in ${\sqs = 13\tev}$ \proton\proton interactions
  measured with the \atlas detector at the \lhc}},
  }{}\href{https://doi.org/10.1140/epjc/s10052-016-4335-y}{Eur.\ Phys.\ J.\
  \textbf{C76} (2016) 502},
  \href{http://arxiv.org/abs/1606.01133}{{\normalfont\ttfamily
  arXiv:1606.01133}}\relax
\mciteBstWouldAddEndPuncttrue
\mciteSetBstMidEndSepPunct{\mcitedefaultmidpunct}
{\mcitedefaultendpunct}{\mcitedefaultseppunct}\relax
\EndOfBibitem
\bibitem{CMS:2011mry}
\cms collaboration, S.~Chatrchyan {\em et~al.},
  \ifthenelse{\boolean{articletitles}}{\emph{{Charged particle transverse
  momentum spectra in pp collisions at ${\sqs = 0.9}$ and $7\tev$}},
  }{}\href{https://doi.org/10.1007/JHEP08(2011)086}{JHEP \textbf{08} (2011)
  086}, \href{http://arxiv.org/abs/1104.3547}{{\normalfont\ttfamily
  arXiv:1104.3547}}\relax
\mciteBstWouldAddEndPuncttrue
\mciteSetBstMidEndSepPunct{\mcitedefaultmidpunct}
{\mcitedefaultendpunct}{\mcitedefaultseppunct}\relax
\EndOfBibitem
\bibitem{CMS:2014kix}
\cms and TOTEM collaborations, S.~Chatrchyan {\em et~al.},
  \ifthenelse{\boolean{articletitles}}{\emph{{Measurement of pseudorapidity
  distributions of charged particles in proton--proton collisions at ${\sqs =
  8\tev}$ by the \cms and TOTEM experiments}},
  }{}\href{https://doi.org/10.1140/epjc/s10052-014-3053-6}{Eur.\ Phys.\ J.\
  \textbf{C74} (2014) 3053},
  \href{http://arxiv.org/abs/1405.0722}{{\normalfont\ttfamily
  arXiv:1405.0722}}\relax
\mciteBstWouldAddEndPuncttrue
\mciteSetBstMidEndSepPunct{\mcitedefaultmidpunct}
{\mcitedefaultendpunct}{\mcitedefaultseppunct}\relax
\EndOfBibitem
\bibitem{CMS:2015zrm}
\cms collaboration, V.~Khachatryan {\em et~al.},
  \ifthenelse{\boolean{articletitles}}{\emph{{Pseudorapidity distribution of
  charged hadrons in proton--proton collisions at ${\sqs = 13\tev}$}},
  }{}\href{https://doi.org/10.1016/j.physletb.2015.10.004}{Phys.\ Lett.\
  \textbf{B751} (2015) 143},
  \href{http://arxiv.org/abs/1507.05915}{{\normalfont\ttfamily
  arXiv:1507.05915}}\relax
\mciteBstWouldAddEndPuncttrue
\mciteSetBstMidEndSepPunct{\mcitedefaultmidpunct}
{\mcitedefaultendpunct}{\mcitedefaultseppunct}\relax
\EndOfBibitem
\bibitem{CMS:2017dou}
\cms collaboration, A.~M. Sirunyan {\em et~al.},
  \ifthenelse{\boolean{articletitles}}{\emph{{Measurement of the inclusive
  energy spectrum in the very forward direction in proton-proton collisions at
  ${\sqs = 13\tev}$}}, }{}\href{https://doi.org/10.1007/JHEP08(2017)046}{JHEP
  \textbf{08} (2017) 046},
  \href{http://arxiv.org/abs/1701.08695}{{\normalfont\ttfamily
  arXiv:1701.08695}}\relax
\mciteBstWouldAddEndPuncttrue
\mciteSetBstMidEndSepPunct{\mcitedefaultmidpunct}
{\mcitedefaultendpunct}{\mcitedefaultseppunct}\relax
\EndOfBibitem
\bibitem{CMS:2017eoq}
\cms collaboration, A.~M. Sirunyan {\em et~al.},
  \ifthenelse{\boolean{articletitles}}{\emph{{Measurement of charged pion,
  kaon, and proton production in proton-proton collisions at ${\sqs =
  13\tev}$}}, }{}\href{https://doi.org/10.1103/PhysRevD.96.112003}{Phys.\ Rev.\
   \textbf{D96} (2017) 112003},
  \href{http://arxiv.org/abs/1706.10194}{{\normalfont\ttfamily
  arXiv:1706.10194}}\relax
\mciteBstWouldAddEndPuncttrue
\mciteSetBstMidEndSepPunct{\mcitedefaultmidpunct}
{\mcitedefaultendpunct}{\mcitedefaultseppunct}\relax
\EndOfBibitem
\bibitem{CMS:2018nhd}
\cms collaboration, A.~M. Sirunyan {\em et~al.},
  \ifthenelse{\boolean{articletitles}}{\emph{{Measurement of charged particle
  spectra in minimum-bias events from proton--proton collisions at ${\sqs =
  13\tev}$}}, }{}\href{https://doi.org/10.1140/epjc/s10052-018-6144-y}{Eur.\
  Phys.\ J.\  \textbf{C78} (2018) 697},
  \href{http://arxiv.org/abs/1806.11245}{{\normalfont\ttfamily
  arXiv:1806.11245}}\relax
\mciteBstWouldAddEndPuncttrue
\mciteSetBstMidEndSepPunct{\mcitedefaultmidpunct}
{\mcitedefaultendpunct}{\mcitedefaultseppunct}\relax
\EndOfBibitem
\bibitem{LHCb-PAPER-2013-070}
LHCb collaboration, R.~Aaij {\em et~al.},
  \ifthenelse{\boolean{articletitles}}{\emph{{Measurement of charged particle
  multiplicities and densities in \proton\proton collisions at
  \mbox{$\sqs=$7\tev} in the forward region}},
  }{}\href{https://doi.org/10.1140/epjc/s10052-014-2888-1}{Eur.\ Phys.\ J.\
  \textbf{C74} (2014) 2888},
  \href{http://arxiv.org/abs/1402.4430}{{\normalfont\ttfamily
  arXiv:1402.4430}}\relax
\mciteBstWouldAddEndPuncttrue
\mciteSetBstMidEndSepPunct{\mcitedefaultmidpunct}
{\mcitedefaultendpunct}{\mcitedefaultseppunct}\relax
\EndOfBibitem
\bibitem{ALICE:2015ial}
\alice collaboration, J.~Adam {\em et~al.},
  \ifthenelse{\boolean{articletitles}}{\emph{{Measurement of pion, kaon and
  proton production in proton--proton collisions at ${\sqs = 7\tev}$}},
  }{}\href{https://doi.org/10.1140/epjc/s10052-015-3422-9}{Eur.\ Phys.\ J.\
  \textbf{C75} (2015) 226},
  \href{http://arxiv.org/abs/1504.00024}{{\normalfont\ttfamily
  arXiv:1504.00024}}\relax
\mciteBstWouldAddEndPuncttrue
\mciteSetBstMidEndSepPunct{\mcitedefaultmidpunct}
{\mcitedefaultendpunct}{\mcitedefaultseppunct}\relax
\EndOfBibitem
\bibitem{ALICE:2020jsh}
\alice collaboration, S.~Acharya {\em et~al.},
  \ifthenelse{\boolean{articletitles}}{\emph{{Production of light-flavor
  hadrons in pp collisions at ${\sqs = 7}$ and ${\sqs = 13\tev}$}},
  }{}\href{https://doi.org/10.1140/epjc/s10052-020-08690-5}{Eur.\ Phys.\ J.\
  \textbf{C81} (2021) 256},
  \href{http://arxiv.org/abs/2005.11120}{{\normalfont\ttfamily
  arXiv:2005.11120}}\relax
\mciteBstWouldAddEndPuncttrue
\mciteSetBstMidEndSepPunct{\mcitedefaultmidpunct}
{\mcitedefaultendpunct}{\mcitedefaultseppunct}\relax
\EndOfBibitem
\bibitem{LHCb-PAPER-2011-037}
LHCb collaboration, R.~Aaij {\em et~al.},
  \ifthenelse{\boolean{articletitles}}{\emph{{Measurement of prompt hadron
  production ratios in \proton\proton collisions at $\sqs=0.9$ and 7\tev}},
  }{}\href{https://doi.org/10.1140/epjc/s10052-012-2168-x}{Eur.\ Phys.\ J.\
  \textbf{C72} (2012) 2168},
  \href{http://arxiv.org/abs/1206.5160}{{\normalfont\ttfamily
  arXiv:1206.5160}}\relax
\mciteBstWouldAddEndPuncttrue
\mciteSetBstMidEndSepPunct{\mcitedefaultmidpunct}
{\mcitedefaultendpunct}{\mcitedefaultseppunct}\relax
\EndOfBibitem
\bibitem{LHCf:2012mtr}
LHCf collaboration, O.~Adriani {\em et~al.},
  \ifthenelse{\boolean{articletitles}}{\emph{{Measurement of forward neutral
  pion transverse momentum spectra for ${\sqs = 7\tev}$ proton-proton
  collisions at the \lhc}},
  }{}\href{https://doi.org/10.1103/PhysRevD.86.092001}{Phys.\ Rev.\
  \textbf{D86} (2012) 092001},
  \href{http://arxiv.org/abs/1205.4578}{{\normalfont\ttfamily
  arXiv:1205.4578}}\relax
\mciteBstWouldAddEndPuncttrue
\mciteSetBstMidEndSepPunct{\mcitedefaultmidpunct}
{\mcitedefaultendpunct}{\mcitedefaultseppunct}\relax
\EndOfBibitem
\bibitem{LHCf:2015rcj}
LHCf collaboration, O.~Adriani {\em et~al.},
  \ifthenelse{\boolean{articletitles}}{\emph{{Measurements of longitudinal and
  transverse momentum distributions for neutral pions in the forward-rapidity
  region with the LHCf detector}},
  }{}\href{https://doi.org/10.1103/PhysRevD.94.032007}{Phys.\ Rev.\
  \textbf{D94} (2016) 032007},
  \href{http://arxiv.org/abs/1507.08764}{{\normalfont\ttfamily
  arXiv:1507.08764}}\relax
\mciteBstWouldAddEndPuncttrue
\mciteSetBstMidEndSepPunct{\mcitedefaultmidpunct}
{\mcitedefaultendpunct}{\mcitedefaultseppunct}\relax
\EndOfBibitem
\bibitem{LHCf:2015nel}
LHCf collaboration, O.~Adriani {\em et~al.},
  \ifthenelse{\boolean{articletitles}}{\emph{{Measurement of very forward
  neutron energy spectra for $7\tev$ proton--proton collisions at the Large
  Hadron Collider}},
  }{}\href{https://doi.org/10.1016/j.physletb.2015.09.041}{Phys.\ Lett.\
  \textbf{B750} (2015) 360},
  \href{http://arxiv.org/abs/1503.03505}{{\normalfont\ttfamily
  arXiv:1503.03505}}\relax
\mciteBstWouldAddEndPuncttrue
\mciteSetBstMidEndSepPunct{\mcitedefaultmidpunct}
{\mcitedefaultendpunct}{\mcitedefaultseppunct}\relax
\EndOfBibitem
\bibitem{LHCf:2018gbv}
LHCf collaboration, O.~Adriani {\em et~al.},
  \ifthenelse{\boolean{articletitles}}{\emph{{Measurement of inclusive forward
  neutron production cross section in proton-proton collisions at ${\sqs =
  13\tev}$ with the LHCf Arm2 detector}},
  }{}\href{https://doi.org/10.1007/JHEP11(2018)073}{JHEP \textbf{11} (2018)
  073}, \href{http://arxiv.org/abs/1808.09877}{{\normalfont\ttfamily
  arXiv:1808.09877}}\relax
\mciteBstWouldAddEndPuncttrue
\mciteSetBstMidEndSepPunct{\mcitedefaultmidpunct}
{\mcitedefaultendpunct}{\mcitedefaultseppunct}\relax
\EndOfBibitem
\bibitem{LHCf:2020hjf}
LHCf collaboration, O.~Adriani {\em et~al.},
  \ifthenelse{\boolean{articletitles}}{\emph{{Measurement of energy flow, cross
  section and average inelasticity of forward neutrons produced in ${\sqs =
  13\tev}$ proton-proton collisions with the LHCf Arm2 detector}},
  }{}\href{https://doi.org/10.1007/JHEP07(2020)016}{JHEP \textbf{07} (2020)
  016}, \href{http://arxiv.org/abs/2003.02192}{{\normalfont\ttfamily
  arXiv:2003.02192}}\relax
\mciteBstWouldAddEndPuncttrue
\mciteSetBstMidEndSepPunct{\mcitedefaultmidpunct}
{\mcitedefaultendpunct}{\mcitedefaultseppunct}\relax
\EndOfBibitem
\bibitem{LHCb-PAPER-2012-034}
LHCb collaboration, R.~Aaij {\em et~al.},
  \ifthenelse{\boolean{articletitles}}{\emph{{Measurement of the forward energy
  flow in \proton\proton collisions at $\sqs= $7\tev}},
  }{}\href{https://doi.org/10.1140/epjc/s10052-013-2421-y}{Eur.\ Phys.\ J.\
  \textbf{C73} (2013) 2421},
  \href{http://arxiv.org/abs/1212.4755}{{\normalfont\ttfamily
  arXiv:1212.4755}}\relax
\mciteBstWouldAddEndPuncttrue
\mciteSetBstMidEndSepPunct{\mcitedefaultmidpunct}
{\mcitedefaultendpunct}{\mcitedefaultseppunct}\relax
\EndOfBibitem
\bibitem{LHCb-DP-2008-001}
LHCb collaboration, A.~A. Alves~Jr.\ {\em et~al.},
  \ifthenelse{\boolean{articletitles}}{\emph{{The \lhcb detector at the LHC}},
  }{}\href{https://doi.org/10.1088/1748-0221/3/08/S08005}{JINST \textbf{3}
  (2008) S08005}\relax
\mciteBstWouldAddEndPuncttrue
\mciteSetBstMidEndSepPunct{\mcitedefaultmidpunct}
{\mcitedefaultendpunct}{\mcitedefaultseppunct}\relax
\EndOfBibitem
\bibitem{LHCb-DP-2014-002}
LHCb collaboration, R.~Aaij {\em et~al.},
  \ifthenelse{\boolean{articletitles}}{\emph{{LHCb detector performance}},
  }{}\href{https://doi.org/10.1142/S0217751X15300227}{Int.\ J.\ Mod.\ Phys.\
  \textbf{A30} (2015) 1530022},
  \href{http://arxiv.org/abs/1412.6352}{{\normalfont\ttfamily
  arXiv:1412.6352}}\relax
\mciteBstWouldAddEndPuncttrue
\mciteSetBstMidEndSepPunct{\mcitedefaultmidpunct}
{\mcitedefaultendpunct}{\mcitedefaultseppunct}\relax
\EndOfBibitem
\bibitem{LHCb-PROC-2010-056}
I.~Belyaev {\em et~al.}, \ifthenelse{\boolean{articletitles}}{\emph{{Handling
  of the generation of primary events in Gauss, the LHCb simulation
  framework}}, }{}\href{https://doi.org/10.1088/1742-6596/331/3/032047}{J.\
  Phys.\ Conf.\ Ser.\  \textbf{331} (2011) 032047}\relax
\mciteBstWouldAddEndPuncttrue
\mciteSetBstMidEndSepPunct{\mcitedefaultmidpunct}
{\mcitedefaultendpunct}{\mcitedefaultseppunct}\relax
\EndOfBibitem
\bibitem{Lange:2001uf}
D.~J. Lange, \ifthenelse{\boolean{articletitles}}{\emph{{The EvtGen particle
  decay simulation package}},
  }{}\href{https://doi.org/10.1016/S0168-9002(01)00089-4}{Nucl.\ Instrum.\
  Meth.\  \textbf{A462} (2001) 152}\relax
\mciteBstWouldAddEndPuncttrue
\mciteSetBstMidEndSepPunct{\mcitedefaultmidpunct}
{\mcitedefaultendpunct}{\mcitedefaultseppunct}\relax
\EndOfBibitem
\bibitem{davidson2015photos}
N.~Davidson, T.~Przedzinski, and Z.~Was,
  \ifthenelse{\boolean{articletitles}}{\emph{{PHOTOS interface in C++:
  Technical and physics documentation}},
  }{}\href{https://doi.org/https://doi.org/10.1016/j.cpc.2015.09.013}{Comp.\
  Phys.\ Comm.\  \textbf{199} (2016) 86},
  \href{http://arxiv.org/abs/1011.0937}{{\normalfont\ttfamily
  arXiv:1011.0937}}\relax
\mciteBstWouldAddEndPuncttrue
\mciteSetBstMidEndSepPunct{\mcitedefaultmidpunct}
{\mcitedefaultendpunct}{\mcitedefaultseppunct}\relax
\EndOfBibitem
\bibitem{Allison:2006ve}
Geant4 collaboration, J.~Allison {\em et~al.},
  \ifthenelse{\boolean{articletitles}}{\emph{{Geant4 developments and
  applications}}, }{}\href{https://doi.org/10.1109/TNS.2006.869826}{IEEE
  Trans.\ Nucl.\ Sci.\  \textbf{53} (2006) 270}\relax
\mciteBstWouldAddEndPuncttrue
\mciteSetBstMidEndSepPunct{\mcitedefaultmidpunct}
{\mcitedefaultendpunct}{\mcitedefaultseppunct}\relax
\EndOfBibitem
\bibitem{Agostinelli:2002hh}
Geant4 collaboration, S.~Agostinelli {\em et~al.},
  \ifthenelse{\boolean{articletitles}}{\emph{{Geant4: A simulation toolkit}},
  }{}\href{https://doi.org/10.1016/S0168-9002(03)01368-8}{Nucl.\ Instrum.\
  Meth.\  \textbf{A506} (2003) 250}\relax
\mciteBstWouldAddEndPuncttrue
\mciteSetBstMidEndSepPunct{\mcitedefaultmidpunct}
{\mcitedefaultendpunct}{\mcitedefaultseppunct}\relax
\EndOfBibitem
\bibitem{LHCb-PROC-2011-006}
M.~Clemencic {\em et~al.}, \ifthenelse{\boolean{articletitles}}{\emph{{The
  \lhcb simulation application, Gauss: Design, evolution and experience}},
  }{}\href{https://doi.org/10.1088/1742-6596/331/3/032023}{J.\ Phys.\ Conf.\
  Ser.\  \textbf{331} (2011) 032023}\relax
\mciteBstWouldAddEndPuncttrue
\mciteSetBstMidEndSepPunct{\mcitedefaultmidpunct}
{\mcitedefaultendpunct}{\mcitedefaultseppunct}\relax
\EndOfBibitem
\bibitem{DeCian:2255039}
M.~De~Cian, S.~Farry, P.~Seyfert, and S.~Stahl,
  \ifthenelse{\boolean{articletitles}}{\emph{{Fast neural-net based fake track
  rejection in the LHCb reconstruction}}, }{}
  \href{http://cdsweb.cern.ch/search?p=LHCb-PUB-2017-011&f=reportnumber&action_search=Search&c=LHCb+Notes}
  {LHCb-PUB-2017-011}, 2017\relax
\mciteBstWouldAddEndPuncttrue
\mciteSetBstMidEndSepPunct{\mcitedefaultmidpunct}
{\mcitedefaultendpunct}{\mcitedefaultseppunct}\relax
\EndOfBibitem
\bibitem{LHCb-DP-2013-002}
LHCb collaboration, R.~Aaij {\em et~al.},
  \ifthenelse{\boolean{articletitles}}{\emph{{Measurement of the track
  reconstruction efficiency at LHCb}},
  }{}\href{https://doi.org/10.1088/1748-0221/10/02/P02007}{JINST \textbf{10}
  (2015) P02007}, \href{http://arxiv.org/abs/1408.1251}{{\normalfont\ttfamily
  arXiv:1408.1251}}\relax
\mciteBstWouldAddEndPuncttrue
\mciteSetBstMidEndSepPunct{\mcitedefaultmidpunct}
{\mcitedefaultendpunct}{\mcitedefaultseppunct}\relax
\EndOfBibitem
\bibitem{Brun:1997pa}
R.~Brun and F.~Rademakers, \ifthenelse{\boolean{articletitles}}{\emph{{ROOT --
  An object oriented data analysis framework}},
  }{}\href{https://doi.org/10.1016/S0168-9002(97)00048-X}{Nucl.\ Instrum.\
  Meth.\  \textbf{A389} (1997) 81}\relax
\mciteBstWouldAddEndPuncttrue
\mciteSetBstMidEndSepPunct{\mcitedefaultmidpunct}
{\mcitedefaultendpunct}{\mcitedefaultseppunct}\relax
\EndOfBibitem
\bibitem{Corti:2006yx}
G.~Corti {\em et~al.}, \ifthenelse{\boolean{articletitles}}{\emph{{Software for
  the \lhcb experiment}},
  }{}\href{https://doi.org/10.1109/TNS.2006.872627}{IEEE Trans.\ Nucl.\ Sci.\
  \textbf{53} (2006) 1323}\relax
\mciteBstWouldAddEndPuncttrue
\mciteSetBstMidEndSepPunct{\mcitedefaultmidpunct}
{\mcitedefaultendpunct}{\mcitedefaultseppunct}\relax
\EndOfBibitem
\bibitem{Tsaregorodtsev:2010zz}
A.~Tsaregorodtsev {\em et~al.},
  \ifthenelse{\boolean{articletitles}}{\emph{{DIRAC3: The new generation of the
  LHCb grid software}},
  }{}\href{https://doi.org/10.1088/1742-6596/219/6/062029}{J.\ Phys.\ Conf.\
  Ser.\  \textbf{219} (2010) 062029}\relax
\mciteBstWouldAddEndPuncttrue
\mciteSetBstMidEndSepPunct{\mcitedefaultmidpunct}
{\mcitedefaultendpunct}{\mcitedefaultseppunct}\relax
\EndOfBibitem
\bibitem{Hunter:2007ouj}
J.~D. Hunter, \ifthenelse{\boolean{articletitles}}{\emph{{Matplotlib: A 2D
  graphics environment}},
  }{}\href{https://doi.org/10.1109/MCSE.2007.55}{Comput.\ Sci.\ Eng.\
  \textbf{9} (2007) 90}\relax
\mciteBstWouldAddEndPuncttrue
\mciteSetBstMidEndSepPunct{\mcitedefaultmidpunct}
{\mcitedefaultendpunct}{\mcitedefaultseppunct}\relax
\EndOfBibitem
\bibitem{Lam:2020a}
S.~K. Lam {\em et~al.},
  \ifthenelse{\boolean{articletitles}}{\emph{{numba/numba: Version 0.52.0}},
  }{} 2020.
\newblock
  doi:~\href{https://doi.org/10.5281/zenodo.4343231}{10.5281/zenodo.4343231}\relax
\mciteBstWouldAddEndPuncttrue
\mciteSetBstMidEndSepPunct{\mcitedefaultmidpunct}
{\mcitedefaultendpunct}{\mcitedefaultseppunct}\relax
\EndOfBibitem
\bibitem{Harris:2020xlr}
C.~R. Harris {\em et~al.}, \ifthenelse{\boolean{articletitles}}{\emph{{Array
  programming with NumPy}},
  }{}\href{https://doi.org/10.1038/s41586-020-2649-2}{Nature \textbf{585}
  (2020) 357}, \href{http://arxiv.org/abs/2006.10256}{{\normalfont\ttfamily
  arXiv:2006.10256}}\relax
\mciteBstWouldAddEndPuncttrue
\mciteSetBstMidEndSepPunct{\mcitedefaultmidpunct}
{\mcitedefaultendpunct}{\mcitedefaultseppunct}\relax
\EndOfBibitem
\bibitem{Virtanen:2019joe}
P.~Virtanen {\em et~al.}, \ifthenelse{\boolean{articletitles}}{\emph{{SciPy
  1.0: Fundamental algorithms for scientific computing in Python}},
  }{}\href{https://doi.org/10.1038/s41592-019-0686-2}{Nat.\ Methods \textbf{17}
  (2020) 261}, \href{http://arxiv.org/abs/1907.10121}{{\normalfont\ttfamily
  arXiv:1907.10121}}\relax
\mciteBstWouldAddEndPuncttrue
\mciteSetBstMidEndSepPunct{\mcitedefaultmidpunct}
{\mcitedefaultendpunct}{\mcitedefaultseppunct}\relax
\EndOfBibitem
\bibitem{Koester:2012a}
J.~K{\"o}ster and S.~Rahmann,
  \ifthenelse{\boolean{articletitles}}{\emph{{Snakemake---a scalable
  bioinformatics workflow engine}},
  }{}\href{https://doi.org/10.1093/bioinformatics/bts480}{Bioinformatics
  \textbf{28} (2012) 2520}\relax
\mciteBstWouldAddEndPuncttrue
\mciteSetBstMidEndSepPunct{\mcitedefaultmidpunct}
{\mcitedefaultendpunct}{\mcitedefaultseppunct}\relax
\EndOfBibitem
\bibitem{Meurer:2017yhf}
A.~Meurer {\em et~al.}, \ifthenelse{\boolean{articletitles}}{\emph{{SymPy:
  Symbolic computing in Python}},
  }{}\href{https://doi.org/10.7717/peerj-cs.103}{PeerJ Comput.\ Sci.\
  \textbf{3} (2017) e103}\relax
\mciteBstWouldAddEndPuncttrue
\mciteSetBstMidEndSepPunct{\mcitedefaultmidpunct}
{\mcitedefaultendpunct}{\mcitedefaultseppunct}\relax
\EndOfBibitem
\bibitem{Schreiner:2021a}
H.~Schreiner {\em et~al.},
  \ifthenelse{\boolean{articletitles}}{\emph{{scikit-hep/boost-histogram:
  Version 0.12.0}}, }{} 2021.
\newblock
  doi:~\href{https://doi.org/10.5281/zenodo.4476368}{10.5281/zenodo.4476368}\relax
\mciteBstWouldAddEndPuncttrue
\mciteSetBstMidEndSepPunct{\mcitedefaultmidpunct}
{\mcitedefaultendpunct}{\mcitedefaultseppunct}\relax
\EndOfBibitem
\bibitem{Dembinski:2020a}
H.~Dembinski {\em et~al.},
  \ifthenelse{\boolean{articletitles}}{\emph{{scikit-hep/iminuit: v2.0.0}}, }{}
  2020.
\newblock
  doi:~\href{https://doi.org/10.5281/zenodo.4310361}{10.5281/zenodo.4310361}\relax
\mciteBstWouldAddEndPuncttrue
\mciteSetBstMidEndSepPunct{\mcitedefaultmidpunct}
{\mcitedefaultendpunct}{\mcitedefaultseppunct}\relax
\EndOfBibitem
\bibitem{Rodrigues:2020a}
E.~Rodrigues {\em et~al.},
  \ifthenelse{\boolean{articletitles}}{\emph{{scikit-hep/particle: Version
  0.14.0}}, }{} 2020.
\newblock
  doi:~\href{https://doi.org/10.5281/zenodo.4292256}{10.5281/zenodo.4292256}\relax
\mciteBstWouldAddEndPuncttrue
\mciteSetBstMidEndSepPunct{\mcitedefaultmidpunct}
{\mcitedefaultendpunct}{\mcitedefaultseppunct}\relax
\EndOfBibitem
\bibitem{Pivarski:2020a}
J.~Pivarski {\em et~al.},
  \ifthenelse{\boolean{articletitles}}{\emph{{scikit-hep/uproot3: 3.14.2}}, }{}
  2020.
\newblock
  doi:~\href{https://doi.org/10.5281/zenodo.4321705}{10.5281/zenodo.4321705}\relax
\mciteBstWouldAddEndPuncttrue
\mciteSetBstMidEndSepPunct{\mcitedefaultmidpunct}
{\mcitedefaultendpunct}{\mcitedefaultseppunct}\relax
\EndOfBibitem
\bibitem{Rodrigues:2020syo}
E.~Rodrigues {\em et~al.}, \ifthenelse{\boolean{articletitles}}{\emph{{The
  Scikit HEP project -- Overview and prospects}},
  }{}\href{https://doi.org/10.1051/epjconf/202024506028}{EPJ Web Conf.\
  \textbf{245} (2020) 06028},
  \href{http://arxiv.org/abs/2007.03577}{{\normalfont\ttfamily
  arXiv:2007.03577}}\relax
\mciteBstWouldAddEndPuncttrue
\mciteSetBstMidEndSepPunct{\mcitedefaultmidpunct}
{\mcitedefaultendpunct}{\mcitedefaultseppunct}\relax
\EndOfBibitem
\bibitem{Needham:2007iz}
M.~Needham and T.~Ruf, \ifthenelse{\boolean{articletitles}}{\emph{{Estimation
  of the material budget of the \lhcb detector}}, }{}
  {\href{https://cds.cern.ch/record/1023537}{CERN-LHCb-2007-025}}, 2007\relax
\mciteBstWouldAddEndPuncttrue
\mciteSetBstMidEndSepPunct{\mcitedefaultmidpunct}
{\mcitedefaultendpunct}{\mcitedefaultseppunct}\relax
\EndOfBibitem
\bibitem{LHCb-DP-2018-002}
M.~Alexander {\em et~al.}, \ifthenelse{\boolean{articletitles}}{\emph{{Mapping
  the material in the LHCb vertex locator using secondary hadronic
  interactions}},
  }{}\href{https://doi.org/10.1088/1748-0221/13/06/P06008}{JINST \textbf{13}
  (2018) P06008}, \href{http://arxiv.org/abs/1803.07466}{{\normalfont\ttfamily
  arXiv:1803.07466}}\relax
\mciteBstWouldAddEndPuncttrue
\mciteSetBstMidEndSepPunct{\mcitedefaultmidpunct}
{\mcitedefaultendpunct}{\mcitedefaultseppunct}\relax
\EndOfBibitem
\bibitem{Barlow:2002yb}
R.~Barlow, \ifthenelse{\boolean{articletitles}}{\emph{{Systematic errors: Facts
  and fictions}},
  }{}\href{http://arxiv.org/abs/hep-ex/0207026}{{\normalfont\ttfamily
  arXiv:hep-ex/0207026}}\relax
\mciteBstWouldAddEndPuncttrue
\mciteSetBstMidEndSepPunct{\mcitedefaultmidpunct}
{\mcitedefaultendpunct}{\mcitedefaultseppunct}\relax
\EndOfBibitem
\bibitem{Barlow:1990vc}
R.~Barlow, \ifthenelse{\boolean{articletitles}}{\emph{{Extended maximum
  likelihood}}, }{}\href{https://doi.org/10.1016/0168-9002(90)91334-8}{Nucl.\
  Instrum.\ Meth.\  \textbf{A297} (1990) 496}\relax
\mciteBstWouldAddEndPuncttrue
\mciteSetBstMidEndSepPunct{\mcitedefaultmidpunct}
{\mcitedefaultendpunct}{\mcitedefaultseppunct}\relax
\EndOfBibitem
\bibitem{Bohm:2013gla}
G.~Bohm and G.~Zech, \ifthenelse{\boolean{articletitles}}{\emph{{Statistics of
  weighted Poisson events and its applications}},
  }{}\href{https://doi.org/10.1016/j.nima.2014.02.021}{Nucl.\ Instrum.\ Meth.\
  \textbf{A748} (2014) 1},
  \href{http://arxiv.org/abs/1309.1287}{{\normalfont\ttfamily
  arXiv:1309.1287}}\relax
\mciteBstWouldAddEndPuncttrue
\mciteSetBstMidEndSepPunct{\mcitedefaultmidpunct}
{\mcitedefaultendpunct}{\mcitedefaultseppunct}\relax
\EndOfBibitem
\bibitem{Fritsch:1980a}
F.~N. Fritsch and R.~E. Carlson,
  \ifthenelse{\boolean{articletitles}}{\emph{{Monotone piecewise cubic
  interpolation}}, }{}\href{https://doi.org/10.1137/0717021}{SIAM J.\ Numer.\
  Anal.\  \textbf{17} (1980) 238}\relax
\mciteBstWouldAddEndPuncttrue
\mciteSetBstMidEndSepPunct{\mcitedefaultmidpunct}
{\mcitedefaultendpunct}{\mcitedefaultseppunct}\relax
\EndOfBibitem
\bibitem{Aaboud:2016mmw}
\atlas collaboration, M.~Aaboud {\em et~al.},
  \ifthenelse{\boolean{articletitles}}{\emph{{Measurement of the inelastic
  proton-proton cross section at ${\sqs = 13\tev}$ with the \atlas detector at
  the \lhc}}, }{}\href{https://doi.org/10.1103/PhysRevLett.117.182002}{Phys.\
  Rev.\ Lett.\  \textbf{117} (2016) 182002},
  \href{http://arxiv.org/abs/1606.02625}{{\normalfont\ttfamily
  arXiv:1606.02625}}\relax
\mciteBstWouldAddEndPuncttrue
\mciteSetBstMidEndSepPunct{\mcitedefaultmidpunct}
{\mcitedefaultendpunct}{\mcitedefaultseppunct}\relax
\EndOfBibitem
\bibitem{LHCb-PAPER-2018-003}
LHCb collaboration, R.~Aaij {\em et~al.},
  \ifthenelse{\boolean{articletitles}}{\emph{{Measurement of the inelastic
  \proton\proton cross-section at a centre-of-mass energy of
  \mbox{$\sqs=$13\tev}}},
  }{}\href{https://doi.org/10.1007/JHEP06(2018)100}{JHEP \textbf{06} (2018)
  100}, \href{http://arxiv.org/abs/1803.10974}{{\normalfont\ttfamily
  arXiv:1803.10974}}\relax
\mciteBstWouldAddEndPuncttrue
\mciteSetBstMidEndSepPunct{\mcitedefaultmidpunct}
{\mcitedefaultendpunct}{\mcitedefaultseppunct}\relax
\EndOfBibitem
\bibitem{Antchev:2017dia}
TOTEM collaboration, G.~Antchev {\em et~al.},
  \ifthenelse{\boolean{articletitles}}{\emph{{First measurement of elastic,
  inelastic and total cross-section at ${\sqs = 13\tev}$ by TOTEM and overview
  of cross-section data at \lhc energies}},
  }{}\href{https://doi.org/10.1140/epjc/s10052-019-6567-0}{Eur.\ Phys.\ J.\
  \textbf{C79} (2019) 103},
  \href{http://arxiv.org/abs/1712.06153}{{\normalfont\ttfamily
  arXiv:1712.06153}}\relax
\mciteBstWouldAddEndPuncttrue
\mciteSetBstMidEndSepPunct{\mcitedefaultmidpunct}
{\mcitedefaultendpunct}{\mcitedefaultseppunct}\relax
\EndOfBibitem
\bibitem{Skands:2014pea}
P.~Skands, S.~Carrazza, and J.~Rojo,
  \ifthenelse{\boolean{articletitles}}{\emph{{Tuning PYTHIA 8.1: The Monash
  2013 tune}}, }{}\href{https://doi.org/10.1140/epjc/s10052-014-3024-y}{Eur.\
  Phys.\ J.\  \textbf{C74} (2014) 3024},
  \href{http://arxiv.org/abs/1404.5630}{{\normalfont\ttfamily
  arXiv:1404.5630}}\relax
\mciteBstWouldAddEndPuncttrue
\mciteSetBstMidEndSepPunct{\mcitedefaultmidpunct}
{\mcitedefaultendpunct}{\mcitedefaultseppunct}\relax
\EndOfBibitem
\bibitem{Ulrich:2021a}
R.~Ulrich, T.~Pierog, and C.~Baus,
  \ifthenelse{\boolean{articletitles}}{\emph{{Cosmic ray Monte Carlo package,
  CRMC}}, }{} 2021.
\newblock
  doi:~\href{https://doi.org/10.5281/zenodo.4558706}{10.5281/zenodo.4558706}\relax
\mciteBstWouldAddEndPuncttrue
\mciteSetBstMidEndSepPunct{\mcitedefaultmidpunct}
{\mcitedefaultendpunct}{\mcitedefaultseppunct}\relax
\EndOfBibitem
\end{mcitethebibliography}

\newpage
\centerline
{\large\bf LHCb collaboration}
\begin
{flushleft}
\small
R.~Aaij$^{32}$,
C.~Abell{\'a}n~Beteta$^{50}$,
T.~Ackernley$^{60}$,
B.~Adeva$^{46}$,
M.~Adinolfi$^{54}$,
H.~Afsharnia$^{9}$,
C.A.~Aidala$^{86}$,
S.~Aiola$^{25}$,
Z.~Ajaltouni$^{9}$,
S.~Akar$^{65}$
% ...

{\footnotesize \it

% ...
$^{9}$Universit{\'e} Clermont Auvergne, CNRS/IN2P3, LPC, Clermont-Ferrand, France\\
% ...
$^{25}$INFN Sezione di Milano, Milano, Italy\\
% ...
$^{32}$Nikhef National Institute for Subatomic Physics, Amsterdam, Netherlands\\
% ...
$^{46}$Instituto Galego de F{\'\i}sica de Altas Enerx{\'\i}as (IGFAE), Universidade de Santiago de Compostela, Santiago de Compostela, Spain\\
% ...
$^{50}$Physik-Institut, Universit{\"a}t Z{\"u}rich, Z{\"u}rich, Switzerland\\
% ...
$^{54}$H.H. Wills Physics Laboratory, University of Bristol, Bristol, United Kingdom\\
% ...
$^{60}$Oliver Lodge Laboratory, University of Liverpool, Liverpool, United Kingdom\\
% ...
$^{65}$University of Cincinnati, Cincinnati, OH, United States\\
% ...
$^{68}$Syracuse University, Syracuse, NY, United States\\
% ...
$^{86}$University of Michigan, Ann Arbor, United States, associated to $^{68}$\\
\bigskip
% ...
\medskip
}
\end{flushleft}


\end{document}